\documentclass{mn2e}
\usepackage{color}
\usepackage{graphicx}
\usepackage{dcolumn}
\usepackage{bm}
\begin{document}
\title[Depleted cores and structural parameter relations]{Depleted
 cores, multi-component fits, and structural parameter relations for luminous early-type galaxies}
\author[B. T. Dullo \& A. W. Graham]
{Bililign T.\ Dullo$^{1}$\thanks{Bdullo@astro.swin.edu.au} and  Alister W.\ Graham$^{1}$  \\
$^1$ Centre for Astrophysics and Supercomputing, Swinburne University
of Technology, Hawthorn, Victoria 3122, Australia.\\}

\maketitle
\label{firstpage}

\begin{abstract}

  New surface brightness profiles from 26 early-type galaxies with
  suspected partially depleted cores have been extracted from the full
  radial extent of {\it Hubble Space Telescope} images. We have
  carefully quantified the radial stellar distributions of the
  elliptical galaxies using the core-S\'ersic model whereas for the
  lenticular galaxies a core-S\'ersic bulge plus an exponential disc
  model gives the best representation. We additionally caution about
  the use of excessive multiple S\'ersic functions for decomposing
  galaxies and compare with past fits in the literature. The
  structural parameters obtained from our fitted models are in
  general, in good agreement with our initial  study using radially
  limited ($R \la 10\arcsec$) profiles, and are used here to update
  several ``central'' as well as ``global'' galaxy scaling
  relations. We find near-linear relations between the break radius
  $R_{b}$ and the spheroid luminosity $L$ such that $R_{b}\propto
  L^{1.13 \pm 0.13}$, and with the supermassive black hole mass
  $M_{\rm BH}$ such that $R_{b}\propto M_{\rm BH}^{0.83 \pm 0.21}$.
  This is internally consistent with the notion that major, dry
  mergers add the stellar and black hole mass in equal proportion,
  i.e., $M_{\rm BH} \propto L$. In addition, we observe a linear
  relation $R_{b}\propto R_{e}^{0.98 \pm 0.15}$ for the core-S\'ersic
  elliptical galaxies---where $R_{e}$ is the galaxies' effective half
  light radii---which is collectively consistent with the
  approximately-linear, bright-end of the curved $L-R_{e}$ relation.
  Finally, we measure accurate stellar mass deficits $M_{\rm def}$
  that are in general 0.5$-4$~$M_{\rm BH}$, and we identify two
  galaxies (NGC 1399, NGC 5061) that, due to their high $M_{\rm
    def}/M_{\rm BH}$ ratio, may have experienced oscillatory
  core-passage by a (gravitational radiation)-kicked black hole.  The
  galaxy scaling relations and stellar mass deficits favor
  core-S\'ersic galaxy formation through a few ``dry'' major merger
  events involving supermassive black holes such that $M_{\rm def}
  \propto M_{\rm BH}^{3.70 \pm 0.76}$, for $M_{\rm BH} \ga 2 \times
  10^{8} M_{\sun}$.

\end{abstract}


\section{Introduction}

Within the hierarchical model of the Universe, galaxy mergers are
thought to be the main formation channel for the build up of massive
elliptical galaxies. These, and fainter, galaxies are known to host a
supermassive black hole (SMBH) at their centre (Magorrian et
al.\ 1998; Richstone et al.\ 1998; Ferrarese \& Ford 2005, and
references therein). It has therefore been hypothesized that the
depleted cores in core-S\'ersic\footnote{A ``core-S\'ersic'' galaxy
  (Graham et al.\ 2003) has a central deficit relative to the inward
  extrapolation of the elliptical galaxy's or bulge's outer S\'ersic
  (1963) profile. As such, it is not equivalent to the ``core'' galaxy
  (Lauer et al.\ 1995) classification. A S\'ersic galaxy has no such
  central deficit.}  galaxies are due to the action of coalescing
black hole binaries which are produced in major, dissipationless (gas
free or ``dry'') mergers of galaxies (e.g., Begelman et al.\ 1980;
Ebisuzaki et al.\ 1991; Milosavljevi\'c \& Merritt 2001; Merritt
2006). Indeed, observations have found binary SMBHs at kpc separation
(e.g., NGC 6240, Komossa et al.\ 2003; Arp 299, Ballo et al.\ 2004;
0402+379, Rodriguez et al.\ 2006; Mrk 463, Bianchi et al.\ 2008), and
slowly more are being found at closer separations (e.g., Burke-Spolaor
2011; Ju et al.\ 2013; Liu et al.\ 2014).  In such
a scenario, three-body interactions involving stars and the SMBH
binary would decay the SMBH binary orbit via the slingshot ejection of
stars from the centres of the ``dry'' galaxy merger remnant, naturally
creating the observed central stellar light (mass) deficits in giant
elliptical galaxies and bulges (e.g., King \& Minkowski 1966; King
1978; Binney \& Mamon 1982; Lauer 1985; Kormendy 1985; Crane et
al.\ 1993; Ferrarese et al.\ 1994, 2006; Lauer et al.\ 1995; Byun et
al.\ 1996; Faber et al.\ 1997; Carollo et al. 1997b).

Numerical simulations targeting the evolution of massive black hole
binaries have predicted that the central stellar mass deficit, $M_{\rm def}$,
of a core-S\'ersic galaxy, i.e., the ejected stellar mass, scales with
the mass of the SMBH binary and the number of (equivalent) major
``dry'' mergers that the galaxy has experienced (e.g., Milosavljevi\'c
\& Merritt 2001; Merritt 2006). Other theoretical studies have
proposed enhanced mass deficits as large as 5 $M_{\rm
  BH}$ as a result of additional stellar ejections from repetitive
core passages of ``recoiled'' SMBHs (e.g., Boylan-Kolchin et al.\ 2004;
Gualandris \& Merritt 2008), or due to the actions of multiple SMBHs
from merging galaxies (Kulkarni \& Loeb 2012).

Tight scaling relations involving the structural parameters of both
core-S\'ersic and S\'ersic galaxies have been shown to exist (e.g.,
Graham 2013 and references therein).
%
%For
%example, the relation between the core size (break radius) $R_{b}$ and
%core surface brightness $\mu (R =R_{b}) =\mu_{b}$ (e.g., Faber et al.\
%1997; Lauer et al.\ 2007a) which we now understand may arise from the
%self-similar nature of the inner S\'ersic profile of luminous
%spheroids (Dullo \& Graham 2012).
% and that between the half-light radius $R_{e}$
%and the effective surface brightness $\mu (R =R_{e}) =\mu_{e}$ (e.g.,
%Kormendy 1977; La Barbera et al.\ 2003; Graham \& Guz\'man 2003;
%Ferrarese et al.\ 2006).
While these correlations can yield clues to the processes of galaxy
formation and evolution, the reliability of this approach depends on
the robustness of the modelling employed for deriving the structural
parameters.
% Aside from the issue of bulge/disc decomposition,
%, there is
%another matter. Faber et al.\ (1997), using the Nuker model fit
%parameters, reported strong correlations among the central properties
%of their luminous ``core'' galaxies which were predicted to be formed
%from major ``dry'' mergers, yet these galaxies' central and global
%properties were found to be only weakly correlated. Subsequent work by
%Graham et al.\ (2003) revealed that 
Although the parameters of the
Nuker model (Grillmair et al.\ 1994; Kormendy et al.\ 1994; Lauer et
al.\ 1995) are known to vary with the fitted radial extent of the
light profile, due to the fact that the straight outer power-law
profile of the Nuker model fails to capture the curved (S\'ersic)
surface brightness profiles of galaxies beyond their core region
(Graham et al.\ 2003; Dullo \& Graham 2012), a work around has been
suggested. Applying the Nuker model to the light profiles of 120
``core'' galaxies, Lauer et al.\ (2007a) noted that the Nuker model
break radii ($R_{b,\rm Nuk}$) are only roughly correlated with the
galaxy properties. However, they identified a better correlated
parameter $r_{\gamma}$ (the ``cusp radius'') as a measure of the core
size (Carollo et al.\ 1997a; Rest et al.\ 2001). Dullo \& Graham
(2012) subsequently showed that this cusp radius closely matches the
break radius of the core-S\'ersic model which we employ
here\footnote{Trujillo et al.\ (2004), Ferrarese et al.\ (2006),
  Richings et al.\ (2011) and Dullo \& Graham (2012, 2013) revealed
  that the Nuker model break radii ($R_{b,\rm Nuk}$) are typically two
  times bigger than the core-S\'ersic model break radii which are
  defined relative to the inward extrapolation of the outer S\'ersic
  function.}. Fitting this model additionally enables us to determine
a galaxy's global structural parameters such as its luminosity and
half light radius, and to measure the central stellar deficit relative
to the outer S\'ersic profile.

As noted by Graham et al.\ (2003), the issue is not only measuring the
core sizes of core-S\'ersic galaxies, but also the misclassification
of coreless ``S\'ersic'' galaxies as galaxies having partially
depleted cores.  Using the core-S\'ersic model, Dullo \& Graham (2012)
found that 18$\%$ of their sample of 39 galaxies which were previously
alleged to have depleted cores according to the Nuker model were
actually S\'ersic galaxies with no cores. Although Lauer (2012)
subsequently reported that the core identification disagreement
between the Nuker model and the core-S\'ersic model was only at the
level of $10\%$, Dullo \& Graham (2013, their Appendix A.2) revisited
and confirmed the 18\% disagreement. In some cases, additional nuclear
components show up as an excess relative to the outer S\'ersic
profile, yet these components' shallow inner profile resulted in the
Nuker model labelling them as ``core'' galaxies.

In this paper we analyse the 26 suspected core-S\'ersic
elliptical galaxies, presenting new light profiles which cover a large
radial range $R \ga 80 \arcsec$. Our sample selection, data reduction,
and light profile extraction technique are discussed in
Section~\ref{Sec2}. In Section \ref{Sec3} we outline our fitting
analysis, provide our results, and additionally compare them to those
from published works, paying attention to the issue of double, triple
and higher S\'ersic model fits.

In Section \ref{Sec4}, we present updated
structural parameters and scaling relations, including central and
global properties, of core-S\'ersic early-type galaxies. In Section
\ref{Sec5.1} we discuss the connection between the galaxy core size
and the black hole mass. In Section \ref{Sec5.2}, we discuss the
methodology that is applied to derive the stellar mass deficits in the
core-S\'ersic early-type galaxies, and in Section \ref{Sec5.3} we then
compare our mass deficits with past measurements. Finally, Section
\ref{Sec5.4} discusses alternative scenarios which have been
presented in the literature for generating cores in luminous
galaxies. Our main conclusions are summarised in Section \ref{Sec6}.

We include three appendices at the end of this paper. The first
presents the core-S\'ersic model fits for all 26 galaxies together
with a review on two galaxies with complicated structures. Notes on
five suspected lenticular galaxies with a bulge plus disc stellar light
distribution are given in the second appendix, while the third
Appendix provides a comparison between this work and modelling by
others of common light profiles. In so doing we highlight a number 
of issues in the literature today that are important but currently
 poorly recognised.

% In so doing we highlight a number of
%issues in the literature today that are important but currently poorly
%recognised.

\begin{center}
\begin{table} 
\begin {minipage}{91mm}
\caption{Updated global parameters and observational
 details from our sample of 31 core-S\'ersic early-type galaxies}
\label{Tabbb1}
\begin{tabular}{@{}llcccc@{}}
\hline
\hline
Galaxy&Type& $M_{V}$ & D &$\sigma $& HST \\
&&(mag)&(Mpc)&(km s$^{-1}$)&Filters\\
(1)&(2)&(3)&(4)&(5)&(6)\\
\multicolumn{1}{c}{} \\              
\hline                           
NGC 0507$^{ w}$        & S0   &  $-22.56$ &$63.7^{n}$ &306&F555W\\
NGC 0584$^{ w}$         & E$^{d}$     $$  &  $-21.14$ &$19.6^{t}$&206&F555W \\
NGC 0741$^{ w}$         & E      $$ &  $-23.39$&  $72.3^{n}$&291&F555W\\
NGC 1016$^{ w}$         & E    $$  &  $-23.31$ & $88.1^{n}$&302&F555W\\
NGC 1399$^{ A}$       & E      $$ &  $-21.91$&  $19.4^{t}$&342&F475W/F814W\\
NGC 1700$^{ w}$          & E      $$  &  $-22.59$ & $53.0 ^{n}$&239&F555W\\
NGC 2300$^{ w}$         & S0     $$  &   $-21.33$ &  $25.7^{n}$&261&F555W\\
NGC 3379$^{ w}$        & E      $$  &  $-20.88$ & $10.3^{t}$ &209&F555W \\
NGC 3608$^{ w}$         & E      $$  &  $-21.07$ &    $22.3^{t}$&192&F555W\\
NGC 3640$^{ w}$        & E     $$ &  $-21.82$ &$ 26.3^{t}$    &182&F555W\\
NGC 3706$^{ w}$       & E     $$  &  $-22.08$ &$45.2^{n}$ & 270&F555W\\
NGC 3842$^{ w}$        & E      $$ &  $-23.14$ & $ 91.0^{n} $& 314&F555W\\
NGC 4073$^{ w}$        & cD$^{d}$     $$  &  $-23.42$ &  $85.3 ^{n}$&275&F555W\\
NGC 4278$^{ A}$         & E      $$  &  $-20.91$ &    $15.6^{t}$&237&F475W/F850LP\\
NGC 4291$^{ w}$        & E      $$ &  $-20.71$ &$ 25.5 ^{t}$ &285&F555W\\
NGC 4365$^{ A}$        & E     $$ &  $-22.02$ &  $19.9^{t}$  &  256&F475W/F850LP\\
NGC 4382$^{ A}$          & S0      $$  &  $-21.38$& $17.9 ^{t}$  &179&F475W/F850LP\\
NGC 4406$^{ A}$        & E      $$  &  $-22.31$ &   $16.7^{t}$ &235&F475W/F850LP\\
NGC 4472$^{ A}$        & E$^{d}$      $$ &  $-22.68$ &  $15.8^{t} $  &294&F475W/F850LP\\
NGC 4552$^{ A}$         & E$^{d}$     $$ &  $-21.26$ & $14.9^{t}$&253&F475W/F850LP\\
NGC 4589$^{ w}$         & E      $$  &  $-21.04$ &$ 21.4^{t}$ &224&F555W\\
NGC 4649$^{ A}$         & E      $$ &  $-22.34$ &$16.4^{t}$&335&F475W/F850LP\\
NGC 5061$^{ w}$         & E      $$  &  $-22.46$ & $32.6^{n}$ &186&F555W\\
NGC 5419$^{ w}$         & E$^{d}$      $$  &  $-23.43$ &  $59.9^{n}$&351&F555W\\
NGC 5557$^{ w}$          & E      $$  &  $-22.39$ &$46.4^{n}$&253&F555W\\
NGC 5813$^{ w}$         & S0   $$ &  $-22.40$ &$ 31.3^{t}$ &237&F555W\\
NGC 5982$^{ w}$         & E      $$  &  $-22.08$ &$41.8^{n}$ &239&F555W\\
NGC 6849$^{ w}$         & SB0      $$  &  $-22.51$ & $80.5^{n}$& 209&F555W\\ 
NGC 6876$^{ w}$         & E      $$&  $-23.51$ & $54.3^{n}$ & 229&F555W\\
NGC 7619$^{ w}$         & E      $$ &  $-22.81$ & $51.5^{t}$  &323&F555W\\
NGC 7785$^{ w}$         & E      $$  &  $-22.01$ & $47.2 ^{n}$&255&F555W\\
\hline
\end{tabular} 

Notes.---Col. (1) Galaxy name. Instrument: $(w)$ {\it HST} WFPC2;
$(A)$ {\it HST} ACS. Col. (2) Morphological classification from the
NASA/IPAC Extragalactic Database
(NED)\footnote{(http://nedwww.ipac.caltech.edu)} except for NGC 3706
and NGC 5813. We adopt an elliptical morphology for NGC 3706 based on
the results in Dullo \& Graham (2013), while for NGC 5813 we adopt an
S0 morphology based on the fitting analysis as well as the photometric
profile (Section~\ref{Sec3}, Appendices \ref{ApppA} and
\ref{ApppB}). The superscript $d$ shows elliptical galaxies which are
classified as disc galaxies in the literature (Appendix \ref{ApppB}).
Col. (3) Absolute \emph{V}-band (galaxy or bulge) magnitude obtained
from Lauer et al.\ (2007b).  These magnitudes are corrected for
Galactic extinction and $(1+z)^{4}$ surface brightness dimming, and
adjusted using the distances from col. (4). The five bulge magnitude
are additionally corrected for inclination and internal dust
attenuation (Driver et al.\ 2008, their Table 1 and Eqs.~1 and 2).
Sources: ($t$) Tonry et al.\ (2001) after reducing their distance
moduli by 0.06 mag (Blakeslee et al.\ 2002); ($n$) from NED (3K
CMB). Col. (5) Central velocity dispersion from
HyperLeda\footnote{(http://leda.univ-lyon1.fr)} (Paturel et
al.\ 2003).  Col. (6) Filters.
% Program ID: GO-5454 (PI: M.\ Franx); GO-5512 and
%GO-6099 (PI: S.\ Faber); GO-6554
%(PI: JP.\ Brodie); GO-6587 (PI: D.\ Richstone); GO-9401
%(PI: P.\ C\^ot\'e); GO-10835 (PI:
%G.\ Sivakoff); GO-10901 (PI: J.\ Blakeslee).
\end {minipage}
\end{table}
\end{center}

\section{Data}\label{Sec2}
\subsection{Sample selection}\label{data}

%In Dullo \& Graham (2012), we analysed the central stellar ($\la 10
%\arcsec$) distributions from a sample of 39 predominantly bright,
%nearby, early-type galaxies which were taken from Lauer et al.\
%(2005). Lauer et al.\ (2005) had identified all these galaxies as
%`core' galaxies having inner regions which are partially depleted of
%stars. Dullo \& Graham (2012) applied the core-S\'ersic model rather
%than the Nuker model (Kormendy et al.\ 1994) and re-classified seven
%of them as ``S\'ersic'' galaxies with no depleted cores. That is, the
%remaining (39-7=) 32 are `core-S\'ersic' galaxies which posses
%depleted cores as reported by Lauer et al.\ (2005). In Dullo \& Graham
%(2013), we then re-analysed all of the six suspected `core-S\'ersic'
%{\it lenticular} galaxies (NGC 507, NGC 2300, NGC 3607, NGC 3706, NGC
%4382 and NGC 6849)---from the sample of 32 core-S\'ersic
%galaxies---using their extended light profiles, out to
%$\sim$$100\arcsec$. One of these six galaxies (NGC 3706) turned out to
%be an elliptical galaxy.

As noted above, in this paper we have targeted 26 suspected
core-S\'ersic elliptical galaxies from Dullo \& Graham (2012) using
more radially extended light profiles. We additionally use the data
for the five core-S\'ersic galaxies (NGC 507, NGC 2300, NGC 3706, NGC
4382 and NGC 6849) from Dullo \& Graham (2013) excluding NGC 3607
because of its dusty nuclear spiral which affected the recovery of the
structural parameters as detailed there. Updated magnitudes, distances
and velocity dispersions pertaining to this combined sample of 31
core-S\'ersic early-type galaxies are presented in Table \ref{Tabbb1}.

\subsection{Galaxy imaging}  

The {\it HST} images for the 26 core-S\'ersic elliptical galaxies,
observed with the ACS and/or WFPC2 cameras were taken from the public
Hubble Legacy Archive (HLA)\footnote{http://hla.stsci.edu}. Although
both the ACS (plate scale of $0\arcsec.046$) and the WFPC2 (plate
scale of $0\arcsec.049$) cameras have comparable high spatial
resolution, the ACS Wide Field Channel (WFC) has a larger, rhomboidal
field-of-view (FOV) of $202\arcsec \times 202 \arcsec$ compared to
that of the WFPC2 CCD array (160$\arcsec$$\times$160$\arcsec$ L-shaped
FOV). We therefore prefer the ACS/F475W ($\sim$ SDSS $g$) and
ACS/F850LP ($\sim$ SDSS $z$) images when available. For NGC 1399,
these were not available, and we instead used the ACS/F475W ($\sim$
SDSS $g$) and ACS/F814W (similar to the Johnson-Cousins $I$-band)
images. For the other galaxies where the ACS images were not available
we use the WFPC2 images taken in the F555W filter (similar to the
Johnson-Cousins V-band). Table \ref{Tabbb1} provides the observation
summaries including the programs, instruments and filters used for
imaging our sample galaxies.

\begin{figure}
\includegraphics[angle=270,scale=0.9]{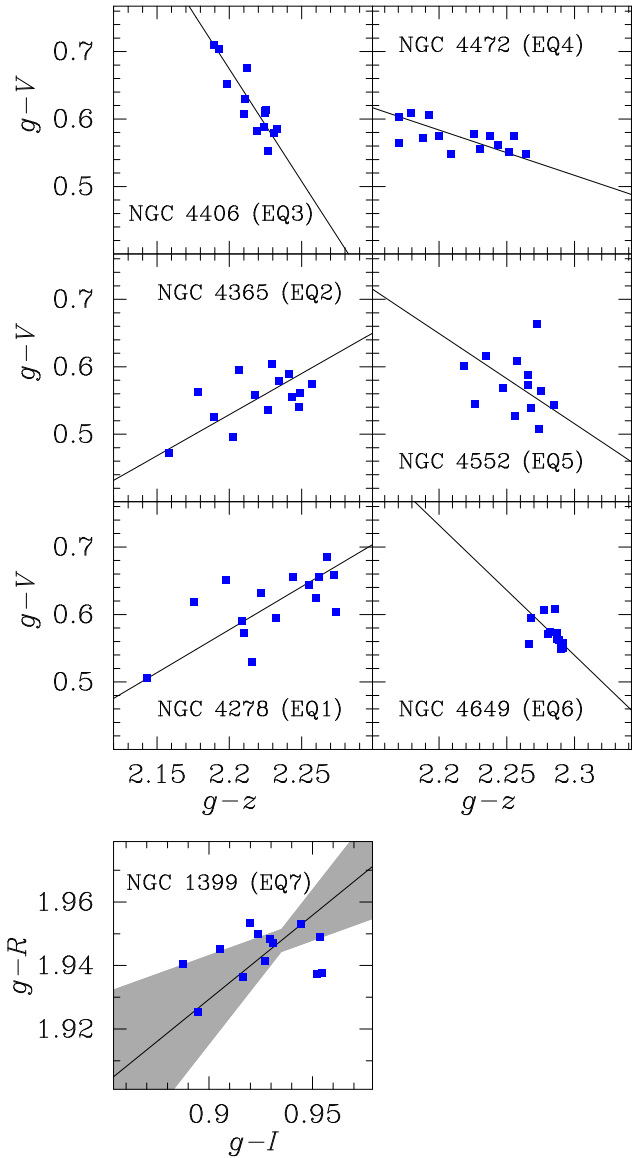}
\caption{Colour transformations for seven galaxies (NGC 1399, NGC
  4278, NGC 4365, NGC 4406, NGC 4472, NGC 4552 and NGC 4649) for which
  the ACS images are available (Table~\ref{Tabbb1}). These colour
  calibrations are from ACS $g$- and $z$-band magnitudes into the
  WFPC2 $V$-band magnitude except for NGC 1399. For NGC 1399, we
  transform the ACS $g$- and $I$-band data into WFPC2 $R$-band data.
  Our $g$-, $z$- and $I$-band data points---each derived here from
  same major-axis with equal position angle and ellipticity using
  the IRAF/{\sc ellipse} fit---along with the Lauer et al.\ (2005)
  $V$- and $R$-band surface brightness profiles are used for creating
  the data points in each panel. Each panel shows the least-squares
  fit and the pertaining equation (EQ) for the galaxy. Shaded regions
  show the 1$\sigma$ uncertainty on the best fits for only one
  galaxy in which the error associated with the slope of the
  least-square fit is large.}
\label{FigIII1}
\end{figure}

\subsection{Surface brightness profiles}\label{Sur_phot} 

As in Dullo \& Graham (2013), we build new, composite light profiles
for all the 26 core-S\'ersic elliptical galaxies by combining the
Lauer et al.\ (2005) very inner ($R\la1\arcsec$) deconvolved F555W
($\sim$ $V$-band) light profiles with our new (calibrated $V$-band)
major-axis light profiles (see below). We have chosen to use the inner
deconvolved light profile from Lauer et al.\ (2005) so that
differences in the core parameters between our works are not
attributed to the use of different treatments of the PSF, but rather
the application of the Nuker versus the core-S\'ersic model. The new
outer profiles are extracted using the IRAF/{\sc ellipse} task
(Jedrzejewski 1987) and cover $\sim100$$\arcsec$ in radius. Our data
reduction steps, along with the surface brightness profile extraction
procedures, are discussed in detail in Sections $2.3$ and $2.4$ of
Dullo \& Graham (2013).

In order to match the ACS $g$ and $z$-band data with the very inner
Lauer et al.\ (2005) deconvolved $V$-band data, we made color
transformations for the six galaxies without $V$-band data (NGC 4278,
NGC 4365, NGC 4406, NGC 4472, NGC 4552 and NGC 4649) using Eqs.\
\ref{EqqIII2}$-$\ref{EqqIII7} which were derived here by applying a
least-squares fit to each galaxy's $(g-V)$ and $(g-z)$ data (see also
Dullo \& Graham 2013 and references therein for a similar practice).
For NGC 1399, Lauer et al.\ (2005) published this galaxy's F606W
(roughly $R$-band) light profile, thus we calibrate our ACS $g$- and
$I$-band data to the $R$-band light profile using Eq.\ \ref{EqqIII1}.
Fig.\ \ref{FigIII1} illustrates these linear fits to the above seven
sample galaxies for which the ACS Wide Field Channel (WFC) images are
available, as listed in Table \ref{Tabbb1}.

\begin{equation}
~~~~~~~~~~~~~~~~~~~~~~ g-V=1.27~(g-z)-2.22.
     \label{EqqIII2}
   \end{equation}

\begin{equation}
~~~~~~~~~~~~~~~~~~~~~~ g-V=1.22~(g-z)-2.15.
     \label{EqqIII3}
   \end{equation}
  
\begin{equation}
~~~~~~~~~~~~~~~~~~~~~~ g-V=-3.33~(g-z)+8.00.
     \label{EqqIII4}
   \end{equation}

\begin{equation}
~~~~~~~~~~~~~~~~~~~~~~ g-V=-0.67~(g-z)+2.06.
     \label{EqqIII5}
   \end{equation}

\begin{equation}
~~~~~~~~~~~~~~~~~~~~~~ g-V=-1.34~(g-z)+3.60.
     \label{EqqIII6}
   \end{equation}
  \begin{equation}
~~~~~~~~~~~~~~~~~~~~~~ g-V=-1.93~(g-z)+4.98.
     \label{EqqIII7}
   \end{equation}

  \begin{equation}
~~~~~~~~~~~~~~~~~~~~~~ g-R=0.53~(g-I)+1.45.
     \label{EqqIII1}
   \end{equation}

\subsubsection{Sky background determination}

Accurate sky level subtraction is of critical importance when
determining the surface brightness profile of a galaxy at large radii.
The galaxy flux is often just a few percent of the sky background
values in the outermost parts. The automatic HLA reduction pipeline
subtracts the sky values from the images. Thus, poor sky subtraction
is a concern  for galaxies which extend beyond the
{\it HST} WFPC2 and ACS FOVs. However, reliable sky background
determination can be done for galaxies with $\mu_{B}=25$ mag
arcsec$^{-2}$ diameters $\la 4 \arcmin$. Not only do the bulk of these
galaxies lie within the WFPC2 and ACS FOVs but their ``counts'' at the
edges of the WFPC2/ACS CCDs are about 10\% fainter than the typical
{\it HST} $V$-band sky value $\sim 22.5$ mag arcsec$^{-2}$ (Lauer et
al.\ 2005). Eight of the 31 sample galaxies (NGC 1399, NGC 3379, NGC
4365, NGC 4382, NGC 4406, NGC 4472, NGC 4552 and NGC 4649) have
major-axis diameter (at $\mu_{B}=25$ mag arcsec$^{-2}$) $\ga 5\arcmin$
(Graham et al.\ 1998; Smith et al.\ 2000; Trager et al.\ 2000).
Fortunately, all these galaxies except two (NGC 1399 and NGC 3379)
have published composite (ACS plus ground-based) data (Ferrarese et
al.\ 2006; Kormendy et al.\ 2009).  The ground-based data enabled
these authors to better constrain the sky level. For NGC 3379,
Schombert \& Smith (2012) published an extended ($\sim 5\arcmin$)
ground-based profile and for NGC 1399, Li et al.\ (2011) published a
very extended ground-based light profile. Our {\it HST}-derived light
profiles for all eight extended galaxies are in a good agreement with
those past published profiles, suggesting only a small or negligible
sky subtraction error by the pipeline.

For the remaining galaxies, we additionally checked the pipeline sky
subtraction by measuring the sky values at the edges of the WFPC2 and
ACS chips , i.e., away from the galaxy and free of contaminating
sources. As expected, the average of the median of the sky values from
several 10 $\times$ 10 pixel boxes is very close to zero for most
galaxies. For a few galaxies, we find that the average values are
slightly below zero even if these galaxies are well within the FOV of
the WFPC2 or ACS. Therefore, we adjust the background level to zero.

\section{Fitting analysis}\label{Sec3}
\subsection{Models for core-S\'ersic elliptical galaxies}\label{Sec3.1} 

In general, the three-parameter S\'ersic (1968) $R^{1/n}$ model, a
generalization of the de Vaucouleurs (1948) $R^{1/4}$ model, is known
to provide accurate representations to the stellar light distributions
of both elliptical galaxies and the bulges of disc galaxies over a
large luminosity and radial range (e.g., Caon et al.\ 1993; D'Onofrio
et al.\ 1994; Young \& Currie 1994; Andredakis et al.\ 1995; Graham et
al.\ 1996). However, because the $R^{1/4}$ model was very popular,
even referred to by many as the $R^{1/4}$ law, Saglia et al.\ (1997)
attempted to explain the observed $R^{1/n}$ light profiles as the sum
of $R^{1/4}$ models and exponential discs.  While this approach had
some merit, in that pressure-supported elliptical galaxies are
becoming increasingly rare, and in fact many intermediate luminosity
early-type galaxies possess rotating discs (e.g., Graham et al.\ 1998;
Emsellem et al.\ 2011; Scott et al.\ 2014), we now know that
lenticular galaxies are very well described by an $R^{1/n}$ bulge plus
an exponential disc which can often additionally contain a bar and/or
a lens (e.g.\ Laurikainen et al.\ 2013, and references therein). The
S\'ersic model's radial intensity distribution can be written as

\begin{equation}
I(R) = I_{e} \exp \left[ - b_{n}
\left(\frac{R}{R_{e}}\right)^{1/n}-1\right],
 \label{EqIII8}
\end{equation}
where $ I_{e}$ is the intensity at the half-light radius $R_{e}$. The
variable $b_{n}\approx 2n- \frac{1}{3}$, for $1\la n\la 10$ (e.g.,\
Caon et al.\ 1993), is coupled to the S\'ersic index $n$, and ensures
that the half-light radius encloses half of the total luminosity. The
luminosity of the S\'ersic model within any radius $R$ is given by

\begin{equation}
L_{T,Ser}(<R) = I_{e} R^{2}_{e} 2 \pi n \frac{e^{b_{n}}}{(b_{n})^{2n}} \gamma (2n,x),
 \label{Eqq9}
\end{equation}
where $\gamma (2n,x)$ is the incomplete gamma function and $x=
b_{n}(R/R_{e})^{1/n}$. The review by Graham \& Driver (2005) 
describes the S\'ersic model in greater detail.

The S\'ersic model fits the surface brightness profiles of the low-
and intermediate-luminosity ($M_{V}\ga-21.5$ mag) spheroids all the
way to the very inner region.  Although, additional nuclear components
are often present in these galaxies and require their own model (e.g.,
Graham \& Guzm\'an 2003; C\^ot\'e et al.\ 2006; den Brok et
al.\ 2014). On the other hand, the inner light profiles of luminous
($M_{V}\la -21.5$ mag) spheroids deviate downward from the inward
extrapolation of their outer S\'ersic model fits (Graham et al.\ 2003;
Trujillo et al.\ 2004; Ferrarese et al.\ 2006).
%Such shallow nuclear profiles of ``core-S\'ersic''
%galaxies are interpreted as an observable signature produced via the
%gravitational slingshot evacuation of stars by decaying binary
%supermassive black holes in ``dry'' major merging events
%(Milosavljevi\'c \& Merritt 2001; Merritt 2006; Faber et al.\ 1997).
In order to describe such galaxies' light distributions, Graham et
al.\ (2003) introduced the core-S\'ersic model which is a combination
of an inner power-law with an outer S\'ersic model. This six-parameter
model is defined as

\begin{equation}
I(R) =I' \left[1+\left(\frac{R_{b}}{R}\right)^{\alpha}\right]^{\gamma /\alpha}
\exp \left[-b\left(\frac{R^{\alpha}+R^{\alpha}_{b}}{R_{e}^{\alpha}}
\right)^{1/(\alpha n)}\right], 
\label{Eqq10}
 \end{equation}
with 
\begin{equation}
I^{\prime} = I_{b}2^{-\gamma /\alpha} \exp 
\left[b (2^{1/\alpha } R_{b}/R_{e})^{1/n}\right].
\label{Eqq11}
\end{equation}
$I_{b}$ is intensity measured at the core break radius $ R_{b}$,
$\gamma$ is the slope of the inner power-law profile, and $\alpha$
controls the sharpness of the transition between the inner power-law
and the outer S\'ersic profile. $R_{e}$ represents the half-light
radius of the outer S\'ersic model, and the quantity $b$ has the same
meaning as in the S\'ersic model (Eq.~\ref{EqIII8}). The total
core-S\'ersic model luminosity (Trujillo et al.\ 2004; their Eq.~A19)
is

$L_{T,cS}=2\pi I'n(R_{e}/b^{n})^{2}\int\limits_{b(R_{b}/R_{e})^{1/n}}^{+\infty}e^{-x}x^{n(\gamma+\alpha)-1}$

\begin{equation}
~~~~~~~~~~~~~~~~~~~~~~~~~~~~~~~~~\times\left[x^{n\alpha}-(b^{n}R_{b}/R_{e})^{\alpha}\right]^{(2-\gamma-\alpha)/\alpha}dx.~~~~~~~
\label{Eqq37}
 \end{equation}

\subsection{Application}\label{Sec3.2}
Fig.~\ref{FigA1} shows our core-S\'ersic model fit to the underlying
host galaxy, major-axis, light distributions for all 26 galaxies. The
fit residuals together with their root-mean-square (rms) values are
given for each galaxy in Appendix \ref{ApppA}.  We find that the light
profile for one of these suspected elliptical galaxies (NGC 5813; see
Appendices \ref{ApppA} and \ref{ApppB}) is better described with a
core-S\'ersic bulge plus an exponential disc model, suggesting an S0
morphology as discussed later. We note that this galaxy's large-scale
disc had negligible contribution to the Dullo \& Graham (2012)
$\sim$$10\arcsec$ light profile fit. Further, in agreement with Dullo
\& Graham (2012), we also detect additional nuclear light components
(i.e., AGN or nuclear star clusters) on the top of the underlying
core-S\'ersic light distributions in six galaxies (NGC 741, NGC 4278,
NGC 4365, NGC 4472, NGC 4552, and NGC 5419). We account for these
nuclear light excesses using a Gaussian function.

Dullo \& Graham (2013) presented the core-S\'ersic(+exponential) model
fits, along with the fit parameters, to five galaxies (NGC 507, NGC
2300, NGC 3706, NGC 4382 and NGC 6849). Table \ref{Tabbb2} presents
the best fit structural parameters for the full (26+5=)31 galaxy
sample obtained by applying our adopted models to the $V$-band,
major-axis, light profiles probing large radial ranges ($R \ga
80\arcsec$).
  
In general, from Appendix~\ref{ApppA} it is apparent that the main
body of luminous ellipticals can be very well described with the
core-S\'ersic model. Our fits yield a median root-mean-square (rms)
residual of $0.045$ mag arcsec$^{-2}$. Out of the full galaxy sample,
only two elliptical galaxies (NGC 4073 and NGC 6876) reveal
complicated structures, as such their light profiles, discussed in
Appendix \ref{ApppA}, are somewhat poorly matched by the core-S\'ersic
model.

% They also
%revealed that its velocity dispersion drops steadily from $\sim230$ km
%s$^{-1}$ at $R=12$$\arcsec$ to $\sim154$ km s$^{-1}$ at $R=
%85\arcsec$.

\begin{figure}
\includegraphics[angle=270,scale=0.56]{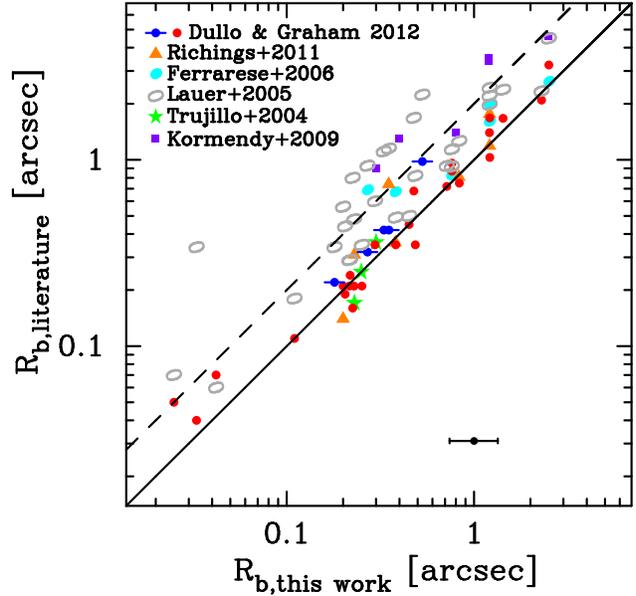}
\caption{Comparison of our major-axis core-S\'ersic break radii
  ($R_{b,\rm this~work}$, Table 2, $V$-band) with previously published
  break radii measurements ($R_{b,\rm literature}$) from (i) Trujillo
  et al.\ (2004, $R$-band, stars), (ii) Lauer et al.\ (2005, $V$-band,
  open ellipses), Ferrarese et al.\ (2006, g-band, blue ellipses),
  Richings et al.\ (2011, $I$- and $H$-bands, triangles) and Dullo \&
  Graham (2012, $V$-band, red circles for ellipticals and blue disc
  symbols for S0s). We also compare the Kormendy et al.\ (2009) inner
  most S\'ersic model fitting radii ($R_{\rm min}$) and our break
  radii for six common galaxies (squares).The geometric-mean break
  radii from Ferrarese et al.\ (2006) were converted into semi-major
  axis break radii using their published galaxy ellipticities at
  $R_{b}$.  The Lauer et al.\ (2005) break radii came from their Nuker
  model fits, while all remaining works mentioned above applied the
  core-S\'ersic model for measuring the galaxies' break radii. The
  break radii from each model is the radius where each model has it's
  maximum curvature. i.e.  where the second derivative of the model's
  intensity profile is a maximum. The solid line indicates a
  one-to-one relation while the dashed line is the $R_{b,\rm
    literature}$ = 2~$R_{b,\rm this~work}$ relation. A representative
  error bar is shown at the bottom of the panel.}
\label{FigIII4}
\end{figure}
\subsubsection{NGC 1700, NGC 3640 and NGC 7785} 

For three sample elliptical galaxies NGC 1700, NGC 3640 and NGC 7785,
while their light profiles are well fit by the core-S\'ersic model
with reasonable rms residuals $ \la 0.063$ mag arcsec$^{-2}$, we find
that their cores are unusually small for their brightnesses. The
core-S\'ersic model yields break radii of $ R_{b} = 0\arcsec.04,
0\arcsec.03,$ and $0\arcsec.03,$ for NGC 1700, NGC 3640, and NGC 7785,
respectively.  Not surprisingly, in Section \ref{Sec4}, it can be seen
that these three questionable cores are outliers in several galaxy
scaling relations involving $R_{b}$, $\sigma$, $M_{\rm BH}$, and
$R_{e}$ and as such they have not been included in the regression
analysis.  For the interested reader, these galaxies are discussed in
more detail in Appendix~\ref{ApppQes}.

%  Interestingly, two (NGC 1700 and NGC 3640) of these galaxies
% have been reported to exhibit features indicative of some recent
% merging activities in the literature (e.g., Franx et al.\ 1989;
% Statler et al.\ 1996; Prugniel et al.\ 1988; Tal et al.\ 2009).
% Section \ref{Sec4}, Section \ref{Sec5} and in particular
% Appendix~\ref{ApppB} provide additional discussions and notes on these
% galaxies.\\

\setlength{\tabcolsep}{0.084in}
\begin{table*}
\begin {minipage}{167mm}
~~~~~~~~~~~\caption{Structural parameters for core-S\'ersic early-type galaxies}
\label{Tabbb2}
\begin{tabular}{@{}lllcccccccccccc@{}}
\hline
\hline
Galaxy&Type&$ \mu_{0, V}$&$ \mu_{b, V} $ & $R _{b}$ &$R_{b}$ &$ \gamma$&$\alpha$&$n$&$R_{e}$&$R_{e}$&$m_{pt, V}$&$\mu_{0d, V}$&$h$\\
&& &&(arcsec)&(pc)&&&&(arcsec)&(kpc)&(mag)&&(arcsec)\\
(1)&(2)&(3)&(4)&(5)&(6)&(7)&(8)&(9)&(10)&(11)&(12)&(13)&(14)\\
\multicolumn{6}{c}{} \\ 
\hline                                           
NGC 0507$^{*}$ &S0&16.16  &16.38      &0.33   &102    &0.07   &5  &2.2   &5.3        &1.65         &     &21.03    &27.69    \\
NGC 0584   &E$^{d}$&13.81      &14.61      &0.21   &21     &0.47   &5  &6.6   &112.5      &11.25        &                    \\
NGC 0741  &E&16.83       &17.52      &0.76   &267    &0.19   &5  &7.4   &53.0       &18.60        &22.1                  \\
NGC 1016 &E&16.35       &17.00      &0.48    &204   &0.15   &2  &5.2   &41.7       &17.79                            \\
NGC 1399$^{+}$&E&15.45   &16.36      &2.30    &202   &0.11   &2  &5.6   &36.6       &3.22                            \\
NGC 1700?  &E&13.34     &13.38      &0.04    &11    &0.19   &5  &6.1   &32.0       &8.23                              \\
NGC 2300$^{*}$&S0& 16.23  &16.61      &0.53    &70    &0.08   &2  &2.2   &7.7        &1.02        &     &20.39   &21.08 \\
NGC 3379  &E&14.76   &15.76      &1.21    &102    &0.19   &2  &5.9  &50.2       &4.21                            \\
NGC 3608 &E&14.56        &15.14      &0.23    &24    &0.29   &5  &6.4   &68.7       &7.28                            \\
NGC 3640?  &E&14.80     &14.72      &0.03    &4     &-0.01  &5  &3.5   &28.0       &2.99                             \\
NGC 3706$^{*}$&E& 14.15       &14.16      &0.11    &24    &-0.02  &10 &6.4   &42.1       &9.18        &                \\
NGC 3842   &E&16.73      &17.42      &0.72    &315   &0.19   &5  &6.9   &102.4      &45.17                      \\
NGC 4073  &cD$^{d}$&16.51       &16.46      &0.22    &90    &-0.06  &10 &6.1   &141.6      &58.61       &       & \\
NGC 4278 &E&15.07        &15.84      &0.83    &52    &0.22   &5  &3.8   &20.2       &1.25        &19.4   &   \\
NGC 4291 &E&14.87         &15.14      &0.30    &36    &0.10   &5  &4.4   &13.6       &1.64         \\
NGC 4365 &E&16.14        &16.50      &1.21    &127   &0.00   &2  &4.8   &47.3       &4.97        &20.2 \\
NGC 4382$^{*}$&S0&14.82   &15.01      &0.27    &24    &0.07   &5  &2.7   &11.1       &0.99        &      &19.50  &35.07     \\
NGC 4406 &E&15.86 &15.97      &0.76    &61    &0.00   &5  &5.5   &145.2      &11.62             \\
NGC 4472 &E$^{d}$&16.18         &16.34      &1.21    &108   &-0.02   &2  &3.0  &48.8       &4.34        &22.0       \\
NGC 4552 &E$^{d}$&14.91        &15.03      &0.38    &17   &0.03   &10 &4.4   &29.7       &1.31        &20.6   \\
NGC 4589&E&14.80         &15.33      &0.20    &27    &0.30   &5  &5.6   &70.8       &9.56          \\
NGC 4649&E&15.75         &16.70      &2.51    &241   &0.21   &2  &3.6   &62.8       &6.02           \\
NGC 5061&E&13.70         &14.09      &0.22    &34    &0.16   &5  &8.4   &68.44      &10.81       \\
NGC 5419&E$^{d}$&17.35       &17.53      &1.43    &416   &-0.06  &2  &5.6   &55.0       &16.01       &19.9  \\
NGC 5557&E&15.05        &15.46      &0.23    &51    &0.19   &5  &4.6   &30.2       &6.80          \\
NGC 5813 &S0&16.04      &16.11      &0.35    &51    &-0.10  &2  &2.8   &7.1        &1.02        &  &20.30   &31.28   \\
NGC 5982&E&15.22         &15.48      &0.25    &51    &0.09   &5  &4.3   &26.8       &5.45                    \\
NGC 6849$^{*}$ &SB0&16.33  &16.67      &0.18    &69    &0.20   &5  &3.2  &7.8         &2.98        &   &20.72  &16.93 \\
NGC 6876&E&17.00        &16.98      &0.45    &119   &0.00  &10 &5.9  & 250.0&65.8  \\
NGC 7619&E&15.41         &15.93      &0.49    &109    &0.16   &5  &7.2   &72.2       &16.23         \\
NGC 7785?&E&14.98         &14.76      &0.03    &5    &0.00   &10 &4.9   &55.1       &12.63 \\
\hline
\end{tabular} 

Notes.---Structural parameters from fits to the $V$-band major-axis surface
brightness profiles (Appendix~\ref{ApppA}). 
% mag arcsec$^{-2}$
The superscript + indicates that we use an $R$-band surface brightness profile instead of a $V$-band surface brightness profile for NGC 1399. The superscript * shows (4S0s and 1E) galaxies for which the fit parameters are taken from Dullo \& Graham (2013). A ``?'' is used to indicate three galaxies with questionable core sizes.  
Col. (1) Galaxy
name. Col. (2) adopted morphological classification. The superscript $d$ shows elliptical galaxies which are classified as disc galaxies in the literature (Appendix \ref{ApppB}). Col. (3)-(11) Best-fit parameters from the core-S\'ersic model, Eq.~\ref{Eqq10}. Col. (12) Central point source apparent
magnitude. Col. (13) Disc central surface brightness. Col. (14) Disc scale length. The surface brightnesses $ \mu_{0}$, $ \mu_{b}$ and $ \mu_{0,d}$ are in units of mag arcsec$^{-2}$.

\end{minipage}
\end{table*}

\begin{figure}
\includegraphics[angle=270,scale=0.56]{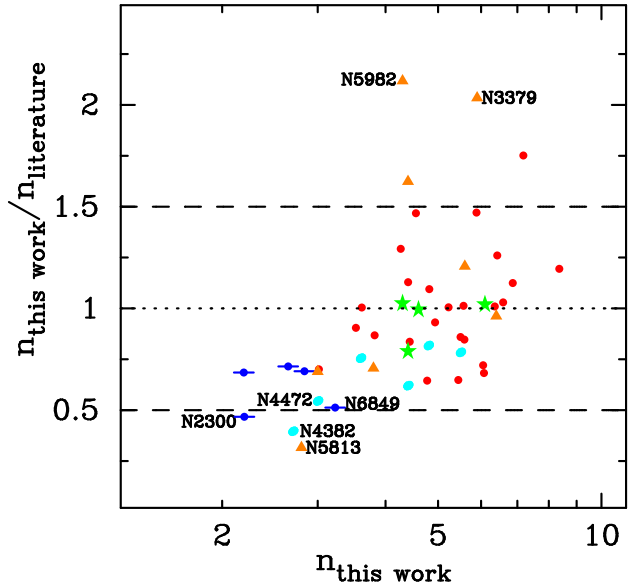}
\caption{Comparison of our major-axis core-S\'ersic galaxy S\'ersic
  index ($n_{\rm this~work}$) with previous S\'ersic values ($n_{\rm
    literature}$) from (i) Trujillo et al.\ (2004, stars), (ii)
  Ferrarese at al.\ (2006, blue ellipses), Richings et al.\ (2011,
  triangles) and Dullo \& Graham (2012, red circles for ellipticals
  and blue disc symbols for S0s). 68\% of the data resides within
  -30\% and +25\% of perfect agreement, and the outliers are explained
  and accounted for in Section 3.3. }
\label{FigIII5}
\end{figure}

\subsection{Comparison with core-S\'ersic fits from the literature}\label{Sec3.1}

Here we illustrate two diagrams comparing our values of $R_{b}$ and $n$ with
those from similar studies in the literature. The agreement is
generally good. We have gone to some effort to identify and explain
all notable disagreements with past studies.

We have three core-S\'ersic galaxies (NGC 4291, NGC 5557, and NGC
5982) in common with Trujillo et al.\ (2004) who used a radial extent
of $\sim$$100\arcsec$. NGC 1700 is also an overlapping galaxy but its
light profile used by Trujillo et al.\ (2004) extends from $R\sim$
$0\arcsec.1$ to $70\arcsec.0$, thus they did not detect the
questionably small core ($R_{b}=0\arcsec.04$) that we potentially
measure here.  With the exception of NGC 1700, classified as a
S\'ersic galaxy by Trujillo et al.\ (2004), there is an excellent
agreement between our break radii and those from Trujillo et
al.\ (2004), see Fig.~\ref{FigIII4}. Their S\'ersic indices are also
consistent for all four galaxies in common with our study, i.e.
including NGC 1700 (Fig.~\ref{FigIII5}).

There are six core-S\'ersic galaxies (NGC 4365, NGC 4382, 4406, NGC
4472, NGC 4552 and 4649) in common with Ferrarese et al.\ (2006) who
used a radial extent of $\sim$$100\arcsec$. For these galaxies, the
Ferrarese et al.\ (2006) geometric-mean, $g$-band, break radii were
taken and converted to major-axis values.  We prefer their g-band than
the z-band data as it more closely matches our $V$-band data. The
agreement between their break radii and our measurements are good
except for three galaxies (NGC 4382, NGC 4552, and NGC 4472). The most
discrepant (by more than 100\%) is the S0 galaxy NGC 4382 but this is
because it was modelled with a core-S\'ersic+exponential model by
Dullo \& Graham (2013, their Fig.\ 3). Ferrarese et al.\ (2006)
treated this galaxy as a single component system, and thus fit the
bulge+disc light with just a core-S\'ersic model, resulting in a
systematically higher $R_{b}$ and $n$ value (Figs.~\ref{FigIII4} and
\ref{FigIII5}, see also Dullo \& Graham 2012, their Fig.\ 7). The
core-S\'ersic stellar light distributions of the remaining two
elliptical galaxies\footnote{We note that NGC 4552 and NGC 4472 were
  classified as S0s by Laurikainen et al.\ (2011), see
  Appendix~\ref{ApppB}.} NGC 4552 and NGC 4472 have a broad and an
intermediate (inner core)-to-(outer S\'ersic) transition region,
respectively, which are well described by the core-S\'ersic model
$\alpha=2$ and $\alpha=5$ values (Fig.~\ref{FigA1}).  This in contrast
to Ferrarese et al's.\ sharp transition ($\alpha=\infty$)
core-S\'ersic model fits which poorly match these two galaxies'
transition regions as can be seen by the systematic bump in their fit
residuals. Due to this, we find a 70\% discrepancy between the
Ferrarese et al.\ (2006) break radii and ours for NGC 4552 and NGC
4472, much bigger than the $1\sigma$ uncertainty range of 10\% quoted
in Dullo \& Graham (2012). However, Dullo \& Graham (2012, their
Fig.\ 7) already discussed the source of this discrepancy for NGC
4552.

Omitting the S0 galaxy NGC 4382, the agreement between the S\'ersic
indices of Ferrrarese et al.\ (2006) and ours is generally good. We
note that $68\%$ of the data in Fig.~\ref{FigIII5} have a S\'ersic
index ratio within -30\% and 25\% of perfect agreement.  This is in
fair agreement with typical uncertainties of 25\% reported for the
S\'ersic index (e.g., Caon et al.\ 1993) and is slightly better than
the Allen et al.\ (2006) $1\sigma$ uncertainty range of $\pm$ 36\%. It
should be remembered that using different filters as well as modelling
minor, major and geometric-mean axis profiles can yield different $n$
values for a galaxy (e.g., Caon et al.\ 1993; Ferrari et al.\ 2004;
Kelvin et al.\ 2012).

 %  This in contrast to Ferrarese et al's.\
% sharp transition ($\alpha=\infty$) core-S\'ersic model fits which
% poorly match these two galaxies' transition regions as can be seen by
% the systematic bump in their fit residuals.  There is about a 70\%
% discrepancy between the Ferrarese et al.\ (2006) break radii and ours
% for these two galaxies, much bigger than the $1\sigma$ uncertainty
% range of 10\% quoted in Dullo \& Graham (2012). This issue was
% demonstrated by Dullo \& Graham (2012, their Fig.\ 7) for NGC 4552.
% Omitting the S0 galaxy NGC 4382, the agreement between the S\'ersic
% indices from these two works is generally good, with $68\%$ of the
% data in Fig.~\ref{FigIII5} having a S\'ersic index ratio within -30\% and
% 25\% of perfect agreement. This is in fair agreement with typical
% uncertainties of 25\% reported for the S\'ersic index (e.g., Caon et al.\
% 1993) and is slightly better than the Allen et al.\ (2006) $1\sigma$
% uncertainty range of $\pm$ 36\%. It should be remembered that using
% different filters as well as modelling minor, major and geometric-mean
% axis profiles may yield different $n$ values for a galaxy (e.g., Caon
% et al.\ 1993; Ferrari et al.\ 2004; Kelvin et al.\ 2012).

\begin{figure}
\includegraphics[angle=270,scale=0.66]{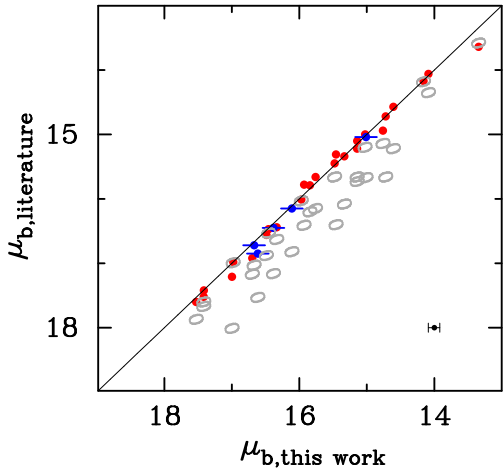}
\caption{Comparison of the break surface brightness measurements from
  this work $\mu_{b,\rm this~work}$ with past measurements $\mu_{b,\rm
    literature}$ from (i) Dullo \& Graham (2012, filled circles and
  the disc symbols are for their core-S\'ersic ellipticals and S0s,
  respectively) and (ii) Lauer et al.\ (2007a, ellipses). These
  surface brightnesses are where the respective intensity profile
  models have the maximum value of their second derivative, i.e. where
  the curvature of the models is greatest.}
\label{FigIII5a}
\end{figure}

We have six core-S\'ersic galaxies (NGC 3379, NGC 3608, NGC 4278, NGC
4472, NGC 4552 and NGC 5813) in common with Richings et al.\ (2011)
whose data extended to $\sim20\arcsec$.  Their break radii for these
common galaxies agree with ours except for NGC 5813. NGC 5813 is a
similar case to that of the S0 galaxy NGC 4382 noted above; it has a
core-S\'ersic bulge+exponential disc light distribution which was
modelled using only the core-S\'ersic model by Richings et
al.\ (2011), thus they measured larger $R_{b}$ and $n$ values. Our
S\'ersic indices agree within 25\% with Richings et al.\ (2011) for
half of the six core-S\'ersic galaxies in common, but for the
remaining half (NGC 3379, NGC 4552 plus the S0 NGC 5813) there is more
than a 40\% discrepancy. For NGC 3379 and NGC 4552, the origin of this
discrepancy appears to be the $\sim$$10\arcsec$ {\it HST}/NICMOS NIC2
F160W light profiles\footnote{The large ellipticals NGC 3379 and NGC
  4552, each with a ($\mu_{B}=25$ mag arcsec$^{-2}$) major-axis
  diameter of $\sim$$5\arcmin$, extend beyond the NICMOS NIC2 CDD.}
used by Richings et al.\ (2011). Their profiles for these two galaxies
may be too limited in radius for the core-S\'ersic model to capture
the actual galaxy light distributions.
% (ii)
%appropriate sky background determination.  
Finally, we note that the elliptical galaxy NGC 5982 is also in common
with Richings et al.\ (2011) who classified it as a S\'ersic galaxy
based on their $n=2$ S\'ersic model fit to the $\sim$$10\arcsec$
NICMOS NIC2 F160W profile.  It seems that Richings et al.\ (2011)
might have missed the core with their $n=2$ S\'ersic fit to this large
elliptical galaxy with $\sigma= 239$ km s$^{-1}$ and $M_{V}=-22.1$
mag.

Kormendy et al.\ (2009) adopted Graham et al's.\ (2003) logic of
defining a core as a deficit in light relative to the inward
extrapolation of a spheroid's outer S\'ersic profile, but they fit the
major-axis light profiles of their core-S\'ersic galaxies using only
the S\'ersic model. They advocate fitting the S\'ersic model over the
radius range where it fits well by eye and distinguishing the core
region in a subjective manner. This exercise assumes no transition
region between the inner core and the outer S\'ersic profile, but
actual galaxy profiles can have a broad transition region. As with the
Nuker model, the ``break radius'' is not the outermost boundary of
this transition region---controlled by the parameter $\alpha$---but
the mid-point of the transition. The outer edge of the transition
region is very hard to judge by eye, and highly subjective. We have
six core-S\'ersic galaxies (NGC 4365, NGC 4382, NGC 4406, NGC 4472,
NGC 4552 and NGC 4649) in common with Kormendy et al.\ (2009).  It is
worth comparing the inner most S\'ersic model fitting radius
($R_{min}$) from Kormendy et al.\ (2009) with our break radius
($R_{b}$) for these six overlapping galaxies: NGC 4365, NGC 4382, NGC
4406, NGC 4472, NGC 4552 and NGC 4649. As shown in Fig.~\ref{FigIII4},
for each core-S\'ersic galaxy in common with Kormendy et al.\ (2009),
their $R_{min}$ are much further out than ours and part of the
explanation likely arises from their method of not measuring the
actual ``break radii''. Fitting the core-S\'ersic model provides this
radius, the extent of the transition region, and the central flux
deficit from within the outer-edge of the transition region.

Figs.~\ref{FigIII4} and \ref{FigIII5a} additionally compare our break
radii and break surface brightnesses, respectively, with those from
Lauer et al.\ (2005) and Dullo \& Graham (2012) for all 31 galaxies
(including the three with questionable core in
Appendix~\ref{ApppQes}). It is important to note that Lauer et
al.\ (2005) fit the Nuker model to their $10\arcsec$ galaxy light
profiles, while Dullo \& Graham (2012) re-modelled these using the
core-S\'ersic model. Our new break radii, determined from spatially
extended ($R \ga 80 \arcsec$) light profiles, are in excellent
agreement with those from Dullo \& Graham (2012), i.e., within the
uncertainty range, except for NGC 1700 and NGC 2300. NGC 1700 is an
elliptical galaxy with a questionably small core mentioned earlier
(see Section \ref{Sec4} and Section \ref{Sec8.1.1} for further
details), while for the S0 galaxy NGC 2300, as noted in Dullo \&
Graham (2013), the contribution of the disc light to the $15\arcsec$
light profile modeled by Dullo \& Graham (2012) resulted in a
bigger break radius and S\'ersic index. Given the remarkable agreement
between the core-S\'ersic break radii of Dullo \& Graham (2012) and
those from their model-independent estimates (their Fig.\ 11), it
implies that our break radii from this work (Table~\ref{Tabbb2}) are
also in a very good agreement with the model-independent radii where
the slope of the logarithmic profile equals -1/2 (Carollo et
al.\ 1997a). In addition, as can be seen in Fig.~\ref{FigIII5a}, our
new break surface brightnesses fully agree with those from Dullo \&
Graham (2012).

% Also,
%Fig.~\ref{FigIII5} shows that the S\'ersic indices from Dullo \&
%Graham (2012) are in a fair agreement with our new values, within 50\%
%except for two S0 galaxies NGC 2300 and NGC 6849 having marginal disc
%light contribution in the Dullo \& Graham (2012) $\sim 15\arcsec$
%%data. The 1$\sigma$ uncertainty on our $n$ in Fig.~\ref{FigIII5} is $\sim
%30\%$.
 
On the other hand, in line with previous core-S\'ersic works, we find
the Nuker model break radii are larger (Fig.~\ref{FigIII4}).  The
Nuker break radii (e.g., Lauer et al.\ 2005, 2007a, b; Krajnovi\'c et
al.\ 2013) are on average two times bigger than our core-S\'ersic
break radii. In estimating larger break radii, the Nuker model
consequently estimates the associated surface brightness up to 2
mag arcsec$^{-2}$ fainter (Fig.~\ref{FigIII5a}).

% More rigorous discussions of the
%differences in the Nuker and core-S\'ersic model fits are given by
%Graham et al.\ (2003), Trujillo et al.\ (2004) and Dullo \& Graham
%(2012) and won't be repeated here.

\begin{figure*}
\includegraphics[angle=270,scale=0.76]{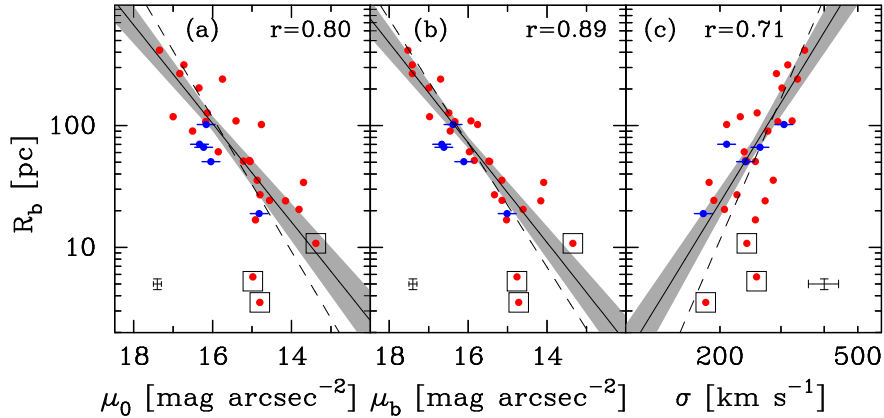}
\caption{Central parameter relations for core-S\'ersic galaxies. The
  core-S\'ersic break radii (Table~\ref{Tabbb2}) are plotted as a
  function of (a) the core-S\'ersic model's central $V$-band surface
  brightness $\mu_{0}$ (of the underlying host galaxy) at the {\it
    HST}/WFPC2 resolution limit $R=0\arcsec.046$, (b) the surface
  brightness $\mu_{b}$ (Table~\ref{Tabbb2}) at the core-S\'ersic
  model's break radius, and (c) the central velocity dispersion $\sigma$
  (Table~\ref{Tabbb1}),
  excluding NGC~1399 for which we have the $R$-band (F606W) profile
  instead of $V$-band profile. Filled circles represent the sample
  elliptical galaxies, while the ``disc'' symbols denote the S0s.
  Three galaxies with questionable cores (NGC 1700, NGC 3640 and NGC
  7785) are enclosed in boxes. The solid lines are the least-squares
  fits to our core-S\'ersic data excluding these three questionable
  galaxies (Table~\ref{Tabbb30}), the shaded regions cover
  the corresponding 1$\sigma$ uncertainties on these regression fits.
  The dashed lines are the least-squares fits to the galaxy data
  including the three questionable galaxies. Pearson correlation
  coefficients, $r$, (and representative error bars) are shown at the
  top (bottom) of each panel.}
\label{FigIII6}
\end{figure*}

%Lastly, we note that the bulges of our S0 galaxies tend to have $n\la
%3$ (Dullo \& Graham 2013), supporting the observation by Balcells et
%al.\ (2003) and Laurikainen et al.\ (2005) that $n=4$ bulges, in
%luminous early-type galaxies, are rare.

\subsection{Multi-component model fits for core-S\'ersic  galaxies}\label{Sec3.4} 

Hopkins et al.\ (2009a, b) fit the surface brightness profiles of both
S\'ersic and core-S\'ersic elliptical galaxies using a double S\'ersic
model. They claimed that these galaxies' outer component is an old
spheroid (with $n \approx 2 -4$) formed by the violent relaxation of
pre-existing stars from a merger event while their inner component was
``excess light'' formed from a dissipative starburst produced by the
same ``wet'' (gas-rich) merger event (e.g., Hernquist et al.\ 1993;
Mihos \& Hernquist 1994).

While this reasonable scenario sounds plausible, we point out two
concerns. First, the lower luminosity early-type (S\'ersic) galaxies
($M_{B} \ga -20.5$ mag) tend to have fast-rotating, outer exponential
discs (Emsellem et al.\ 2011; Krajnovi\'c et al.\ 2013), rather than
old, outer spheroid-like components. Second, the higher luminosity
(core-S\'ersic) elliptical galaxies have a central deficit of light
rather than an excess, and are thought to be formed from dry merger
events (e.g., Faber et al.\ 1997). The violent relaxation simulations
yield $n \approx 2 -4$ (e.g., van Albada 1982; McGlynn 1984) and
therefore cannot account for luminous elliptical galaxies with $n \ga
4$ built from dry mergers. In addition, they do not explain how
low-luminosity elliptical galaxies with $n \la 2$ are made, nor why
these galaxies follow the same $L-n$ relation as the massive
ellipticals with $n \ga 4$.

%We note that the depleted cores
%of core-S\'ersic galaxies do not emerge naturally from the double
%S\'ersic model fits, although Hopkins et al.\ (2009b) argued that
%core-S\'ersic ellipticals are re-merged ``excess light'' galaxies
%where stars are ejected from the centres of their inner excess light
%component by sinking binary SMBHs.

The Hopkins et al.\ (2009a, b) galaxy sample included lenticular
galaxies (NGC 507, NGC 1400, NGC 2778, NGC 4382, NGC 4459, NGC 4476,
NGC 5813 and NGC 4515) which were thought to be ellipticals and
modelled using an inner S\'ersic model plus an outer S\'ersic model
with $n\approx 2.5$. Our fits in Fig.~\ref{FigA1} and those in Dullo
\& Graham (2013) show that NGC 507, NGC 4382 and NGC 5813 are
core-S\'ersic lenticular galaxies that are well described by a
core-S\'ersic bulge plus an exponential disc model with very small rms
residuals of 0.027 mag arcsec$^{-2}$, 0.016 mag arcsec$^{-2}$ and
0.015 mag arcsec$^{-2}$, respectively. Laurikainen et al.\ (2010,
2011) have also detected a weak nuclear bar in the unsharp mask image
of NGC 507. Further, all of our core-S\'ersic elliptical galaxies
(Table~\ref{Tabbb1}), except for NGC 1016, NGC 3706 and NGC 4073, are
in common with Hopkins et al.\ (2009b). As discussed in
Section~\ref{Sec3.2}, these galaxies are well described by the
core-S\'ersic model, which can be seen from the residual
profiles\footnote{The (not shown) residual profile from the
  core-S\'ersic fit to the simulated light profile by Hopkins et
  al.\ (2008, their Fig.~2) would reveal systematic structures that
  have not been properly modelled.  Similarly, the residual profile
  from the double S\'ersic fit to, for example, AM 2246-490 (Hopkins
  et al.\ 2008, their Fig.\ 6) would also reveal telling mismatches.}
(Fig.~\ref{FigA1}) which also have small rms residuals $\sim$ 0.045
mag arcsec$^{-2}$. This can be compared to the larger rms residuals
given by Hopkins et al.\ (2009b, their Figs.~10-14) from their double
S\'ersic model fits for six core-S\'ersic galaxies in common with our
sample (NGC 4365, NGC 4382, NGC 4406, NGC 4472, NGC 4552 and NGC
4649). That is, with better fits, we have shown that these galaxies
have a central deficit of light, in accord with the dry merging
scenario involving supermassive black holes. In stark contrast,
Hopkins et al.\ (2009b) argued that these galaxies support a wet
merger scenario.

% That is, both elliptical and 
%lenticular core-S\'ersic galaxies are better thought of as having a
%central deficit in its bulge, rather than labelling it as an ``excess
%light'' galaxy.

Similarly, Dhar \& Williams (2012) argued that both S\'ersic and
core-S\'ersic galaxies can be represented well by two or three ``DW''
functions which are the 2D projections of the 3D Einasto density model
(Dhar \& Williams 2010). As in Hopkins et al.\ (2009a, b), all their
sample galaxies, including the six common core-S\'ersic galaxies (NGC
4365, NGC 4382, NGC 4406, NGC 4472, NGC 4552 and NGC 4649), are
presented as having an inner ``extra light'' component that has a
half-light radius $R_{e}$ $\sim 0.2 - 1$ kpc. This is somewhat similar
to the ground-based work by Huang et al.\ (2013) who argued that
elliptical galaxies comprise three distinct components---an inner
($R_{e}\la 1$ kpc) component, a middle ($R_{e} \approx 2.5$ kpc)
component and an outer $R_{e}\la 10$ kpc envelope---which are all
represented by S\'ersic models with $n\approx 1- 2$, at odds with the
traditional picture of violent relaxation producing $R^{1/2}- R^{1/4}$
profiles.  In contrast, the fits from our study show that the inner
$\sim$ kpc of “ellipticals” (excluding the depleted core) are not
disconnected from, but are rather the simple extensions of, their
outer regions\footnote{Although, we do note that some luminous cD
  galaxies (e.g., Seigar et al.\ 2007) may have excess flux at large
  radii with respect to their inner S\'ersic profiles as a result of
  minor mergers (Hilz et al.\ 2013) and accretion from their cluster
  environment.}. Of course disturbed, unrelaxed galaxies, especially
those with peculiar morphology, won't be well described by a single
core-S\'ersic model. These particular galaxies may well appear to have
multiple (S\'ersic) spheroidal components.

In addition to a handful of S0 galaxies (IC~2006, NGC~4697), a handful
of barred S0 galaxies (IC 4329, NGC 6673), and a handful of unrelaxed
peculiar galaxies (NGC~2305, NGC~4976), Huang et al.\ (2013) included
15 cD galaxies (IC~1633, IC~2597, IC~4765, IC~4797, NGC~596, NGC~1172,
NGC~1339, NGC~1427, NGC~3087, NGC~4696, NGC~4786, NGC~6909, NGC~6958,
NGC~7192, NGC~7796) in their galaxy sample. The tell-tale signature of
a fit which has failed to fully capture the curvature in the radial
stellar distribution is evidenced by the pattern in the residual
profile.  This can be seen in, for example, Fig.\ 36 from Huang et
al.\ (2013), where, from 150$\arcsec$ to beyond 400$\arcsec$ there is
a systematic hump in their residual profile for ESO 185-G054. The
artificial ring in their residual image also reveals that the fit is
not optimal. This is because the 9 parameters of their three fitted
S\'ersic models have collectively managed to approximate the light
profile out to a radius $R^{1/4}=3.5~(R=150\arcsec)$. In this
instance, the use of a fourth, extended S\'ersic model would have
enabled a better fit to the outer half of the light profile, as in the
case of ES0 221-G026 which Huang et al.\ (2013) fit with 4 S\'ersic
models. However, this does not mean that the galaxy ESO 185-G054
actually has 4 components, simply that if one uses enough parameters
then one can better approximate the light profile. Rather than
applying a multitude of S\'ersic components, we advocate trying to
establish which components are real and then applying the appropriate
function, as done by, for example, L\"asker et al. (2014).

To continue this important point, but avoid creating too much of a
distraction in the main text, in Appendix~\ref{ApppC} we provide
comparisons between our core-S\'ersic modelling of our galaxy sample
and some other recent works which obtained dramatically different
results for galaxies with depleted cores.
%did not use the core-S\'ersic model and obtained different
%results.

%\subsection{Core-S\'ersic lenticular galaxies}

%Dullo \& Graham (2013) found four lenticular galaxies with depleted
%cores. Here, we NGC 5813 is a core-S\'ersic lenticular galaxy. These
%disc galaxies are. Appendix~\ref{ApppC} provides comparisons to some
%recent works which used models other than the core-S\'ersic model.

\section{Structural parameter relations}\label{Sec4} 

We explore several galaxy structural parameter relations for 28
core-S\'ersic early-type galaxies with carefully acquired
core-S\'ersic parameters.

\subsection{The core's size and brightness}\label{Sec4.1}

\begin{figure}
\includegraphics[angle=270,scale=0.76]{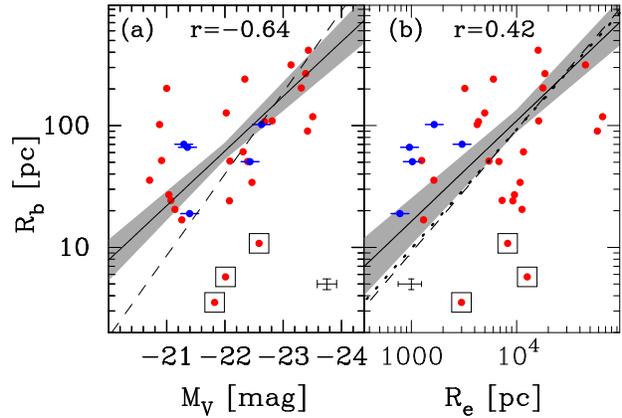}
\caption{Similar to Fig.~\ref{FigIII6}, but shown here are the
  correlations between the core-S\'ersic break radius $R_{b}$ and (a)
  absolute $V$-band magnitude of a galaxy or a bulge for a disc galaxy
  (Table~\ref{Tabbb1}), and (b) effective (half-light) radius $R_{e}$
  (Table~\ref{Tabbb2}). In Fig.~\ref{FigIII7}(b) we also include
  least-squares fit to the $R_{b}$ and $R_{e}$ data for just the
  elliptical galaxies (dotted line).}
\label{FigIII7}
\end{figure}

\begin{figure*}
\includegraphics[angle=270,scale=0.76]{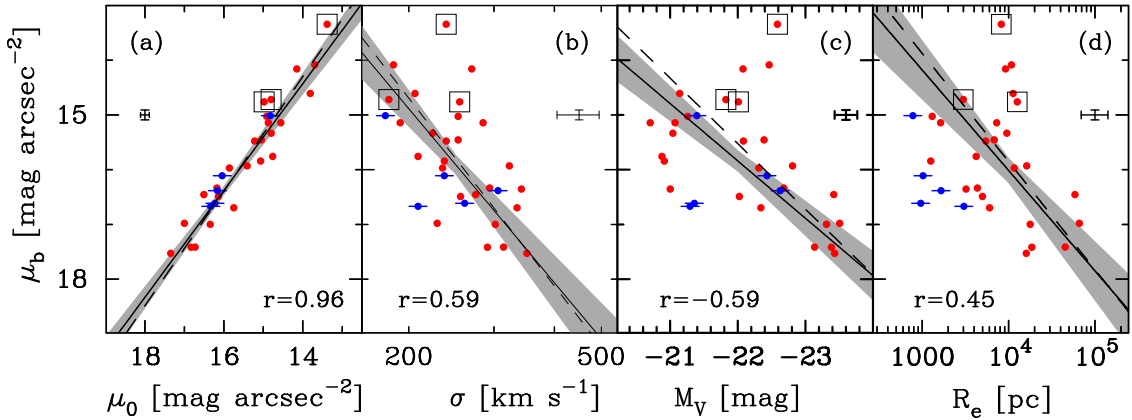}
\caption{Correlations between the core-S\'ersic break surface
  brightness $\mu_{b}$ (Table~\ref{Tabbb2}) and (a) the core-S\'ersic
  model central $V$-band surface brightness of the host galaxy
  $\mu_{0}$ (Table~\ref{Tabbb2}), (b) central velocity dispersion
  $\sigma$ (Table~\ref{Tabbb1}), (c) absolute galaxy (bulge) magnitude
  (Table~\ref{Tabbb1}), and (d) effective (half-light) radius $R_{e}$
  (Table~\ref{Tabbb2}). Symbolic representations are as in
  Fig.~\ref{FigIII6}. The solid lines are the least-squares fits to
  our core-S\'ersic data, the shaded regions cover
  the corresponding 1$\sigma$ uncertainties on these regression fits.
  The dashed lines are the least-squares fits to the galaxy data
  including the questionable galaxies.  Pearson correlation
  coefficients, $r$, (and representative error bars) are shown at the
  bottom (top) of each panel. }
\label{FigIII8}
\end{figure*}

The good agreement between the structural parameters from this work
and those form our initial study using $\sim$$10\arcsec$ profiles
(Dullo \& Graham 2012) suggests that the different correlations that
will be shown here agree with those of Dullo \& Graham
(2012). Fig.~\ref{FigIII6} shows the relation between the
core-S\'ersic break radius $R_{b}$ (Table~\ref{Tabbb2}) and central
galaxy properties including (a) the core-S\'ersic model's central
$V$-band surface brightness $\mu_{0}$ (Table~\ref{Tabbb2}), (b) the
break surface brightness $\mu_{b}$ (Table~\ref{Tabbb2}) and (c) the
velocity dispersion $\sigma$ (Table~\ref{Tabbb1}). The solid and
dashed lines shown in each panel of Figs.~\ref{FigIII6} to
\ref{FigIII9} are two distinct linear regression fits obtained with
and without the three elliptical galaxies (NGC 1700, NGC 3640 and NGC
7785) with questionably small cores ($R_{b} \le 11~\rm{pc}$). We note
that all the relations given in Table~\ref{Tabbb30} are for the galaxy
data without these three questionable galaxies. Using the ordinary
least squares (OLS) bisector regression from Feigelson \& Babu (1992),
a fit to the $R_{b}$ and $\mu_{0}$ data gives $R_{b} \propto
\mu_{0}^{0.40 \pm 0.05}$, while applying the bisector fit to $R_{b}$
and $\mu_{b}$ yields $R_{b} \propto \mu_{b}^{0.41\pm 0.04}$, and the
bisector fit to $R_{b}$ and $\sigma$ yields $R_{b} \propto
\sigma^{4.65 \pm 0.52}$ (Table~\ref{Tabbb30}).
\begin{center}
\begin{table*}
\begin {minipage}{160mm}
~~~~~~~~~~~\caption{Structural parameter relations}
\label{Tabbb30}
~~~~~~~~\begin{tabular}{@{}llccccccccc@{}}
\hline
\hline
Relation&OLS bisector fit&$\Delta$ (Vertical scatter)\\

\multicolumn{2}{c}{} \\ 
\hline  
$R_{b}-\mu_{0}$& $\mbox{log}\left(\frac{R_{b}}{\mbox{pc}}\right)= (0.40\pm 0.05)\left(\mu_{0}-16\right) +~(2.02~ \pm 0.05)$&0.24 dex\\                          
$R_{b}-\mu_{b}$&$\mbox{log}\left(\frac{R_{b}}{\mbox{pc}}\right)= (0.41\pm 0.04)\left(\mu_{b}-16\right) +~(1.86~ \pm 0.04)$&0.18 dex\\

$R_{b}-\sigma$&$\mbox{log}\left(\frac{R_{b}}{\mbox{pc}}\right)= (4.65\pm 0.52)~\mbox{log}\left(\frac{\sigma}{250~\mbox{km s} ^{-1}}\right) +~(1.82~ \pm 0.06)$&0.29 dex\\
$R_{b}-M_{V}$&$\mbox{log}\left(\frac{R_{b}}{\mbox{pc}}\right)= (-0.45\pm 0.05)\left(M_{V}+22\right) +~(1.79~ \pm 0.06)$&0.30 dex\\
$R_{b}-R_{e}$&$\mbox{log}\left(\frac{R_{b}}{\mbox{pc}}\right)= (0.83\pm 0.10)~\mbox{log}\left(\frac{R_{e}}{10^{4}~\rm{pc}}\right) +~(2.04~ \pm 0.08)$&0.43 dex\\
$R_{b}-R_{e}~ \mbox{(for ellipticals only)}$&$\mbox{log}\left(\frac{R_{b}}{\mbox{pc}}\right)= (0.98\pm 0.15)~\mbox{log}\left(\frac{R_{e}}{10^{4}~\rm{pc}}\right) +~(1.97~ \pm 0.10)$&0.45 dex\\
$\mu_{b}-\mu_{0}$&$\mu_{b}= (0.98\pm 0.06)\left(\mu_{0}-16\right) +~(16.38~ \pm 0.06)$&0.28\\
$\mu_{b}-\sigma$& $\mu_{b}= (10.56\pm 1.58)~\mbox{log}\left(\frac{\sigma}{250~\mbox{km s} ^{-1}}\right) +~(15.91~ \pm 0.16)$&0.80\\
$\mu_{b}-M_{V}$&$\mu_{b}= (-1.05\pm 0.13)\left(M_{V}+22\right) +~(15.84~ \pm 0.18)$&0.80\\
$\mu_{b}-R_{e}$&$\mu_{b}= (1.83\pm 0.31)\left(\frac{R_{e}}{10^{4}~\rm{pc}}\right) +~(16.01~ \pm 0.21)$&1.17\\
$R_{b}-M_{\rm BH}$  ($M-\sigma$ derived $M_{\rm BH}$ for 23 galaxies &$\mbox{log}\left(\frac{R_{b}}{\mbox{pc}}\right)= (0.80\pm 0.10)~\mbox{log}\left(\frac{M_{\rm BH}}{10^{9} M_{\sun}}\right) +~(2.01~ \pm 0.05)$&0.27 dex\\
~~~~~~~~~~~~~~~~~~~~~~~~~~~~~plus 8 direct $M_{\rm BH}$ masses)&&\\
$R_{b}-M_{\rm BH}$ ($M-L$ derived $M_{\rm BH}$ for 23 galaxies &$\mbox{log}\left(\frac{R_{b}}{\mbox{pc}}\right)= (0.79\pm 0.08)~\mbox{log}\left(\frac{M_{\rm BH}}{10^{9} M_{\sun}}\right) +~(1.75~ \pm 0.06)$&0.27 dex\\
~~~~~~~~~~~~~~~~~~~~~~~~~~~~~plus 8 direct $M_{\rm BH}$ masses)&&\\
\hline
\end{tabular} 
\end {minipage}
\end{table*}
\end{center}
%\begin{equation} 
%\mbox{log}\left(\frac{R_{b}}{\mbox{pc}}\right)= (0.40\pm 0.05)\left(\mu_{0}-16\right) +~(2.02~ \pm 0.05),~~   
%\label{Eqq12}
%\end{equation}
%while, applying the bisector fit to $R_{b}$ and $\mu_{b}$  yields
%\begin{equation}
%\mbox{log}\left(\frac{R_{b}}{\mbox{pc}}\right)= (0.41\pm 0.04)\left(\mu_{b}-16\right) +~(1.86~ \pm 0.04).~~   
%\label{Eqq13}
%\end{equation}
%The bisector fit to $R_{b}$ and $\sigma$ is
%\begin{equation}
%\mbox{log}\left(\frac{R_{b}}{\mbox{pc}}\right)= (4.65\pm 0.52)~\mbox{log}\left(\frac{\sigma}{250~\mbox{km s} ^{-1}}\right) +~(1.82~ \pm 0.06).~~   
%\label{Eqq14}
%\end{equation}
%The SMBH masses $M_{\rm BH}$ used in Fig.~\ref{FigIII6}(d) are
%acquired from direct SMBH mass measurements for 7 galaxies, while for
%the remaining 23, the BH masses are predicted using the Graham \&
%Scott (2013) core-S\'ersic $M-\sigma$ relation. As such, our
%$R_{b}-M_{BH}$ relation is somewhat dependent on our $R_{b}-\sigma$
%relation. Applying the bisector fit to the $R_{b}$ and $M_{\rm BH}$
%data gives
%\begin{equation}
%\mbox{log}\left(\frac{R_{b}}{\mbox{pc}}\right)= (0.80\pm 0.11)~\mbox{log}\left(\frac{M_{\rm BH}}{10^{9} M_{\sun}}\right) +~(2.02~ \pm 0.06).~~  
%\label{Eqq15} 
%\end{equation}  

Similar to Fig.~\ref{FigIII6}, Fig.~\ref{FigIII7} reveals that the
core-S\'ersic break radii $R_{b}$ correlate with global galaxy
properties such as (a) the $V$-band absolute magnitude $M_{V}$
(Table~\ref{Tabbb1}) and (b) the effective radius $R_{e}$
(Table~\ref{Tabbb2}). The bisector fit gives the near-linear relation
between $R_{b}$ and $M_{V}$ as $R_{b} \propto L^{1.13\pm0.13}$, while
the fitted relation for the $R_{b}$ and $R_{e}$ data is $R_{b} \propto
R_{e}^{0.83\pm 0.10}$ (Table~\ref{Tabbb30}). Of all the relations
(Figs.~\ref{FigIII6} and \ref{FigIII7}, Table~\ref{Tabbb30}), the
weakest correlation with a Pearson correlation coefficient of $r=0.42$
is between $R_{b}$ and $R_{e}$, while the for the remaining relations
$|r|\ge 0.7$.

We note that, intriguingly, the bulges seem to reveal a systematic
trend in the $R_{b}-\mu_{0}$, $R_{b}-\mu_{b}$ (Figs~\ref{FigIII6}a and
\ref{FigIII6}b) and $R_{b}-R_{e}$ (Fig.~\ref{FigIII7}b) diagrams, although it
is more obvious in the $R_{b}-R_{e}$ plane. For a given break radius,
bulges appear to be compact, i.e., $R_{e} \la 2$ kpc (see also Dullo
\& Graham 2013 and Graham 2013) and possess somewhat fainter central
and break surface brightnesses. Thus, in Fig~\ref{FigIII7}(b) we
additionally include the OLS bisector fit to the relation between
$R_{b}$ and $R_{e}$ (dotted line) for only the elliptical galaxies,
which is given by $R_{b} \propto R_{e}^{0.98 \pm 0.15}$
(Table~\ref{Tabbb30}). Combining this relation with the $R_{b} \propto
L^{1.13\pm 0.13}$ relation (Table~\ref{Tabbb30}, Fig.~\ref{FigIII7}a)
gives $R_{e} \propto L^{1.15 \pm 0.22}$ for elliptical galaxies with
$M_{B} \la -20.5$ mag. This can be compared with the bright end of the
curved $L-R_{e}$ relation given in Section 5.3.1 of Graham \& Worley
(2008). The linear regression fit to the luminous ($M_{B} \la -19.0$
mag) galaxies in Graham \& Worley (2008, their Fig.\ 11a) has a slope
of 0.9 (see also Liu et al.\ 2008, Bernardi et al.\ 2007), and it is
$10-25\%$ steeper at brighter luminosities.
%\begin{equation}
%\mbox{log}\left(\frac{R_{b}}{\mbox{pc}}\right)= (0.98\pm 0.15)~\mbox{log}
%\left(\frac{R_{e}}{10^{4}~\rm{pc}}\right) +~(1.97~ \pm 0.10),~~   
%\label{Eqq18}
%\end{equation}
While the elliptical galaxies appear to follow the steeper near-linear
$R_{b}$ $\propto$ $R_{e}^{0.98}$ relation than the $R_{b}$ $\propto$
$R_{e}^{0.83}$ relation for the combined (elliptical+bulge) sample,
the vertical rms scatters for both these relations are large
(Table~\ref{Tabbb30}).

%Within the errors, the near-linear elliptical galaxy $R_{b}-R_{e}$
%relation can be approximated to $R_{b}\approx 0.01
%R_{e}$, which should be compared to the $R_{b}\approx
%0.02^{+0.025}_{-0.01} R_{e}$ result by C\^ot\'e et al.\ (2007).

The tight correlations seen in the $R_{b}-\mu_{b}$ and $R_{b}-\sigma$
diagrams (Figs.~\ref{FigIII6}b and \ref{FigIII6}c) were also shown by
Faber et al.\ (1997, their Fig.\ 8), Lauer et al.\ (2007a, their
Figs.\ 4 and 6) and Dullo \& Graham (2012, their Fig.\ 17). Similar
trends to the $R_{b}-M_{V}$ distribution (Fig.\ref{FigIII7}a) can be
seen in the works by Faber et al.\ (1997, their Fig.~4), Ravindranath
et al.\ (2001, their Fig.\ 5a, b), Laine et al.\ (2003, their
Fig.\ 9), Trujillo et al.\ (2004, their Fig.\ 9), de Ruiter et
al.\ (2005, their Fig.\ 8), Lauer et al.\ (2007a, their Figs.\ 5) and
Dullo \& Graham (2012, their Fig.\ 17). The slope of the $R_{b}-L$
relation $1.13 \pm 0.13$ (Table~\ref{Tabbb30}) that we find here can
be compared to the similar slopes 1.15, 0.72, $1.05 \pm 0.10$, $1.32
\pm 0.11$ and $1.45 \pm 0.23$ published by Faber et al.\ (1997), Laine
et al.\ (2003), de Ruiter et al.\ (2005), Lauer et al.\ (2007a) and
Dullo \& Graham (2012), respectively. It is worth noting that the
$R_{b}-L$ relations in Faber et al.\ (1997), Laine et al.\ (2003) and
de Ruiter et al.\ (2005) were derived using the Nuker break radii,
while Lauer et al.\ (2007a) used the ``cusp radius''---the radius at
which the negative logarithmic slope of the Nuker model equals
0.5. The slopes (not the intercepts) of the $R_{b}-L$ relations
obtained using the core-S\'ersic model and the Nuker model break radii
can coincidentally be consistent because of the way the Nuker model
systematically overestimates the break radius in comparison with the
core-S\'ersic model. Also provided here, achieved using well
constrained core-S\'ersic fit parameters, are the $R_{b}-\mu_{b}$ and
$R_{b}-\sigma$ relations which are consistent with Dullo \& Graham
(2012, their Eqs.~7, and 5) within the errors. Due to coupling of
$R_{b}$ and $\mu_{b}$ along the light profile (Dullo \& Graham 2012,
their Figs.~17c and 18), our $R_{b}-\mu_{b}$ relation agrees with that
of Lauer et al.\ (2007a, their Eq.\ 17).

As shown in Fig.~\ref{FigIII6}(b), the core size of a galaxy ($R_{b}$)
and its surface brightness ($\mu_{b}$) are closely related.
Fig.~\ref{FigIII8} reveals that $\mu_{b}$ is thus also tightly
correlated with (a) the central surface brightness $\mu_{0}$, (b) the
velocity dispersion $\sigma$, (c) the spheroid absolute magnitude
$M_{V}$, and (d) the effective (half-light) radius $R_{e}$.  The OLS
bisector fits are given in Table \ref{Tabbb30}. The $\mu_{b}-\sigma$
and $\mu_{b}-L$ relations (Table~\ref{Tabbb30}) agree with those in
Dullo \& Graham (2012, their Eqs.~10 and 9, respectively).

Lastly, given the disagreement between the ``core'' parameters
($\gamma, R_{b}, \mu_{b}$) of the core-S\'ersic model and the Nuker
model (Section~\ref{Sec3.1}), it is expected that our galaxy scaling
relations (Table~\ref{Tabbb30}) may differ from similar relations
obtained using the Nuker model. However, as mentioned above, some of
the slopes (not the intercepts) of these scaling relations derived
from these two models can agree. Moreover, the close agreement between
the core-S\'ersic break radius and the Lauer et al.\ (2007a) ``cusp
radius'' (Dullo \& Graham 2012) suggests that the scaling relations
based on these two core measurements would be consistent.

\section{The central stellar deficit} \label{Sec5} 
 \subsection{Core size versus black hole mass}\label{Sec5.1} 

\begin{figure}
\includegraphics[angle=270,scale=0.76]{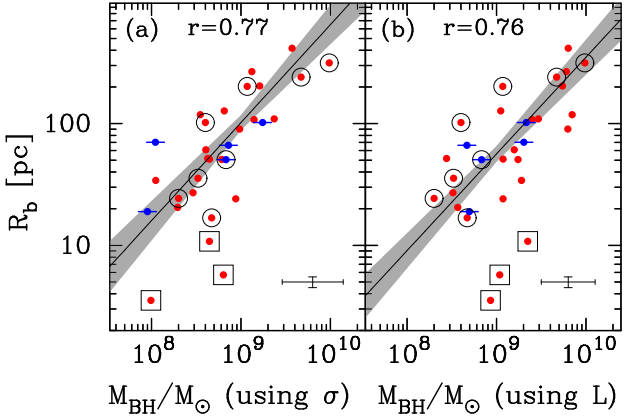} 
\caption{Correlation between core-S\'ersic break radius $R_{b}$
  (Table~\ref{Tabbb2}) and black hole mass. The SMBH masses are
  acquired from direct SMBH mass measurements for 8 circled galaxies,
  while for the remaining 23 galaxies the SMBH masses are predicted
  using either (a) the Graham \& Scott (2013) ``non-barred $M_{\rm
    BH}-\sigma$'' relation (Table~\ref{Tabbb3}) or (b) their $B$-band
  core-S\'ersic $M_{\rm BH}-L$ relation which is converted here to the
  $V$-band using $B-V=1.0$ (Fukugita et al.\ 1995). Symbolic
  representations are as in Fig.~\ref{FigIII6}.  The solid lines are
  the least-squares fits to our core-S\'ersic data, the shaded regions
  cover the corresponding 1$\sigma$ uncertainties on these regression
  fits.  Pearson correlation coefficients, $r$, (and representative
  error bars) are shown at the bottom (top) of each panel. }
\label{FigIII9} 
\end{figure}

 If galaxy core formation proceeds by the orbital decay of black hole
 binaries, from merging galaxies, as suggested by simulations (e.g.,
 Ebisuzaki et al.\ 1991; Merritt 2006) and advocated by Faber et al.\
 (1997), then a close relation between the core size ($R_{b}$) and the
 black hole mass ($M_{\rm BH}$) of a galaxy might be expected.  Given
 the well known $M_{\rm BH}-\sigma$ (Ferrarese \& Merritt 2000;
 Gebhardt et al.\ 2000) and $M_{\rm BH}-L$ (Marconi \& Hunt 2003;
 Graham \& Scott 2013) relations, the strong $R_{b}-\sigma$ and
 $R_{b}-L$ correlations in Section~\ref{Sec4.1} hint at a tight
 $R_{b}-M_{\rm BH}$ relation. This trend is observed in
 Fig.~\ref{FigIII9}, and quantified in Table \ref{Tabbb30} for our
 sample of 31 galaxies. Eight of these galaxies have direct SMBH mass
 measurements, while the remaining SMBH masses were predicted using
 either the Graham \& Scott (2013) ``non-barred $M_{\rm BH}-\sigma$''
 relation\footnote{We are able to use the non-barred $M-\sigma$
   relation, which has smaller uncertainties, because it is consistent
   with the core-S\'ersic $M-\sigma$ relation.}
\begin{equation}
\mbox{log}\left(\frac{M_{\rm BH}}{M_{\sun}}\right)= (5.53\pm 0.34)~\mbox{log}\left(\frac{\sigma}{\rm 200~km s^{-1}}\right) +~(8.22~ \pm 0.05),
\label{Eqq34a}
\end{equation}
 or their $B$-band
 core-S\'ersic $M_{\rm BH}-L$ relation which is converted here to
 $V$-band using $B-V=1.0$ (Fukugita et al.\ 1995; Faber et al.\ 1997),
 to give
\begin{equation}
\mbox{log}\left(\frac{M_{\rm BH}}{M_{\sun}}\right)= (-0.54\pm 0.12)~\left(M_{V}+22\right)+~(9.03~ \pm 0.09).
\label{Eqq34b}
\end{equation} 
Following Graham et al.\ (2011, see the discussion in their Section
2.1.1), we assumed a 10 \% uncertainty on our velocity dispersions in
order to estimate the errors on the SMBH masses which were predicted
using the $M_{\rm BH}-\sigma$ relation. The predicted masses are
given in Table~\ref{Tabbb3}. Note that since the resulting
$R_{b}-M_{\rm BH}$ distributions, shown in Figs.~\ref{FigIII9}a and b,
are primarily driven by galaxies with predicted SMBH masses, the
observed trend may simply be due to the existence of the
$R_{b}-\sigma$ (Fig.~\ref{FigIII6}c), $R_{b}-L$ (Fig.~\ref{FigIII7}a)
relations and the $M_{\rm BH}-\sigma$, $M_{\rm BH}-L$ relations,
although we find below that this is not the case.

\begin{figure}
\includegraphics[angle=270,scale=0.73]{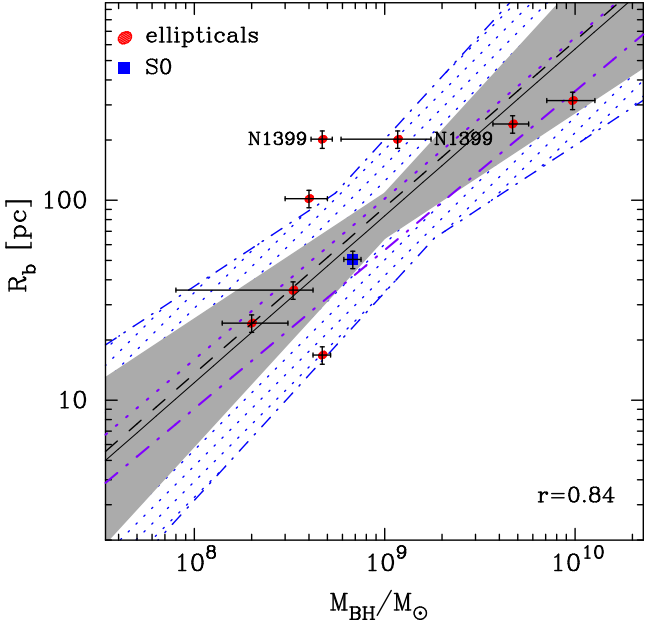}
\caption{Correlation between core-S\'ersic break radius $R_{b}$
  (Table~\ref{Tabbb2}) and black hole mass (Table~\ref{Tabbb3}) for
  eight galaxies with dynamically determined (i.e., direct) black hole
  mass $M_{\rm BH}$ measurements. For NGC 1399 we include two direct
  SMBH mass measurements: (i) (4.7$\pm$0.6)$\times$$10^{8}M_{\sun}$
  (Gebhardt et al.\ 2007) and (ii)
  (1.2$\pm$$0.58$)$\times$$10^{9}$$M_{\sun}$ (Houghton et al.\ 2006).
  The solid line is the least-squares fit assuming the Houghton et
  al.\ (2006) SMBH mass for NGC 1399, while the dashed line uses the
  Gebhardt et al.\ (2007) mass, see the text for further detail. The
  inner shading marks the $1$$\sigma$ uncertainty on Eq.~\ref{Eqq34},
  while the outer shading extends this by 0.26 dex (the rms scatter)
  in the log $M_{\rm BH}$ direction. The dotted and dashed-dotted lines
  are the fits for the full galaxy sample shown in
  Figs.~\ref{FigIII9}(a) and (b), respectively (Table~\ref{Tabbb30}).
  The Pearson correlation coefficient, r, obtained when using the
  Houghton et al.\ (2006) SMBH mass for NGC 1399 is shown at the
  bottom of the panel. }
\label{FigIII11} 
\end{figure}

Fig.~\ref{FigIII11} plots the $R_{b}-M_{\rm BH}$ relation for the
eight galaxies with directly measured SMBH masses. NGC 1399 has two
distinct SMBH mass measurements in the literature (Houghton et al.\
2006, $M_{\rm BH}=1.2^{+0.58}_{-0.58} \times 10^{9} M_{\sun}$;
Gebhardt et al.\ 2007, $M_{\rm BH}=4.7^{+0.60}_{-0.60} \times 10^{8}
M_{\sun}$).  While this galaxy has a normal core (202 pc) for its
absolute magnitude ($M_{V}=-21.9$ mag), it appears that the $M_{\rm
  BH}=4.7^{+0.60}_{-0.60} \times 10^{8} M_{\sun}$ dynamical SMBH mass
measurement of Gebhardt et al.\ (2007) {\it may} be too small for its
202 pc core size. This mass estimate makes the galaxy an obvious
outlier in the $R_{b}-M_{\rm BH}$ diagram (Fig.~\ref{FigIII11}).
Using the Houghton et al.\ (2006) SMBH mass for NGC 1399, the OLS
bisector fit to the $R_{b}$ and $M_{\rm BH}$ data for our eight
galaxies with direct SMBH measurements yields
\begin{equation}
\mbox{log}\left(\frac{R_{b}}{\mbox{pc}}\right)= (0.83\pm 0.21)~\mbox{log}\left(\frac{M_{\rm BH}}{10^{9} M_{\sun}}\right) +~(1.92~ \pm 0.11),~~   
\label{Eqq34}
\end{equation} 
with an rms scatter of $\Delta=0.26$ dex in the log $M_{\rm BH}$
direction.  When using the SMBH mass for NGC 1399 from Gebhardt et
al.\ (2007) rather than from Houghton et al.\ (2006), $R_{b} \propto
M_{\rm BH}^{0.83\pm0.23}$ with an intercept of 1.96 $\pm$ 0.14
(Fig.~\ref{FigIII11}, dashed line). While this relation is in
excellent agreement with Eq.~\ref{Eqq34}, the scatter in this
distribution is larger (0.35 dex in the log $M_{\rm BH}$ direction).
%Aside from NGC 1399, two elliptical galaxies (NGC 3379, NGC 4552) are
%outliers in the $R_{b}-M_{\rm BH}$ diagram (Fig.~\ref{FigIII11}).

We note that, as shown in Fig.~\ref{FigIII11}, the near-linear
$R_{b}-M_{\rm BH}$ relation established by these eight galaxies
(Eq.~\ref{Eqq34}, solid line) is consistent with the relations
constructed by including the remaining sample galaxies with predicted
SMBH mass measurements (Figs.~\ref{FigIII9}a and b,
Table~\ref{Tabbb30}). However, the $M_{\rm BH}-L$ relation
(Eq.~\ref{Eqq34b}, Graham \& Scott 2013) appears to somewhat
overpredict the SMBH masses relative to the $M_{\rm BH}-\sigma$
relation (Eq.~\ref{Eqq34a}, Graham \& Scott 2013) for our
core-S\'ersic galaxy sample. This can be seen from the smaller
intercept of the ($M_{\rm BH}-L$)-based $R_{b}-M_{\rm BH}$ relation,
1.75 $\pm$ 0.06 (Fig.~\ref{FigIII9}b, Table~\ref{Tabbb30}), compared
to the intercept of the ($M_{\rm BH}-\sigma$)-based $R_{b}-M_{\rm BH}$
relation, $2.01 \pm 0.05$ (Fig.~\ref{FigIII9}a, Table~\ref{Tabbb30}).
This unexpected situation has arisen because of a difference in the
$L-\sigma$ relation between the core-S\'ersic galaxy sample in Graham
\& Scott (2013) and that used here. We find a 2.44$\sigma$ difference
between the intercepts of the $L-\sigma$ relations from these two
studies\footnote{We found consistent results using both the linear
  regression code employed in this paper and in Graham \& Scott
  (2013).}, which largely explains the 3.25$\sigma$ difference between
the intercepts of the ($M_{\rm BH}-\sigma$)-based and ($M_{\rm
  BH}-L$)-based $R_{b}-M_{\rm BH}$ relations (Table~\ref{Tabbb30}).

%For our 31 core-S\'ersic galaxies
%(Table~\ref{Tabbb1}), we find a relation 

% \[
% \mbox{log}\left(\sigma\right)= (-0.12 \pm 0.02)\left(M_{V}+22.0\right) +~(2.39~ \pm 0.02),
% \]
% while for the Graham \& Scott (2013) core-S\'ersic
% galaxies we convert their $B$-band magnitudes into $V$-band and obtain
 
% \[
% \mbox{log}\left(\sigma\right)= (-0.12 \pm 0.02)\left(M_{V}+22.0\right) +~(2.44~ \pm 0.02).
% \]

%  Similarly, using
% the Two Micron All Sky Survey (2MASS) $K_{s}$-band absolute
% magnitudes, we find $L_{K_{s}} \propto \sigma^{3.90 \pm 0.68}$ with a
% zero point log ($\sigma$$) =2.39 \pm 0.02$ at $M_{K_{s}}=-25.0$ mag,
% and $L_{K_{s}} \propto \sigma^{4.30 \pm 0.76}$ with a zero point log
% ($\sigma$$) =2.43 \pm 0.02$ at $M_{K_{s}}=-25.0$ mag for our and
% Graham \& Scott (2013) core-S\'ersic galaxies, respectively. Note that
% these $V$- and $K_{s}$-band magnitudes are corrected for Galactic
% extinction and cosmological redshift dimming, and for S0s, we
% additionally corrected for inclination and internal dust attenuation.
% Also, the 2MASS magnitudes are corrected for the flux missed by the
% 2MASS aperture (see Schombert \& Smith 2012) using Eq.~1 (and Fig.~2)
% from Scott, Graham \& Schombert (2013).

% and it is roughly consistent with the ``direct''
%$R_{b}-M_{\rm BH}$ relation, $1.92 \pm 0.11$, (Eq.~\ref{Eqq34}).

To further appreciate the (galaxy core)-(SMBH mass) connection we
derive additional $R_{b}-M_{\rm BH}$ relations by combining the
non-barred $M_{\rm BH}-\sigma$ relation from Graham \& Scott (2013)
with the $R_{b}-\sigma$ relation (Table~\ref{Tabbb30}) to obtain the
new $R_{b}-M_{\rm BH}$ relation

\begin{equation}
\mbox{log}\left(\frac{R_{b}}{\mbox{pc}}\right)= (0.87\pm 0.18)~\mbox{log}\left(\frac{M_{\rm BH}}{10^{9} M_{\sun}}\right) +~(2.06~ \pm 0.17),~~   
\label{Eqq35}
\end{equation}
which is in good agreement with Eq.~\ref{Eqq34}. Similarly,
combining the $R_{b} - L$ relation (Table~\ref{Tabbb30}) with the
core-S\'ersic $M_{\rm BH}-L$ relation from Graham \& Scott (2013,
their Table 3 $B$-band), which is converted
here to the $V$-band using $B-V$=1.0, gives

\begin{equation}
\mbox{log}\left(\frac{R_{b}}{\mbox{pc}}\right)= (0.83\pm 0.21)~\mbox{log}\left(\frac{M_{\rm BH}}{10^{9} M_{\sun}}\right) +~(1.72~ \pm 0.20).~~   
\label{Eqq36}
\end{equation}

Although we only have eight galaxies with direct black hole mass
measurements, Eq.~\ref{Eqq34} is consistent (i.e., overlapping
$1\sigma$ uncertainties) with the two inferred relations
(Eqs.~\ref{Eqq35} and \ref{Eqq36}).

% It is worth noting that
%all the updated relations given here are based on accurate
%core-S\'ersic fit parameters that are determined by modelling new,
%spatially extended light profiles. The good agreement between these
%inferred relations contradicts with the discussion in Lauer et al.\
%(2007a).

\begin{center}
\begin{table*}

\begin {minipage}{145mm}
\caption{Core-S\'ersic galaxy data.}
\label{Tabbb3}
\begin{tabular}{@{}llcccccc@{}}
\hline
\hline
Galaxy&$V-I$&$M/L_{V}$&log~($L_{\rm def}/L_{\sun,V}$)&log~($M_{\rm def}/M_{\sun}$)&log~($M_{\rm BH}/M_{\sun}$)&$M_{\rm def}/M_{\rm BH}$\\
&&&&&&\\
(1)&(2)&(3)&(4)&(5)&(6)&(7)\\
\multicolumn{1}{c}{} \\              
\hline
NGC 0507 &1.40[a] &5.5&8.34&9.08&9.24$\pm0.39$[p]&0.69\\
NGC 0584 &1.27[b]&4.5&7.78&8.43 &8.29$\pm0.39$[p]&1.38\\
NGC 0741 &1.32[a]&5.0&8.96&9.66&9.12$\pm0.39$[p]&3.43\\
NGC 1016 &1.32[a]&5.0&9.10&9.80&9.21$\pm 0.39$[p]&3.85\\
NGC 1399 &1.21[a]&4.0&9.57&9.97&8.67 $^{+0.05}_{-0.06}$, 9.07 $^{+0.17}_{-0.46}$[d]&20.0, 7.90\\
NGC 1700 &1.29[b]&4.8&7.77&8.45&8.65$\pm 0.39$[p]&0.63\\
NGC 2300 &1.33[b]&5.0&7.81&8.51&8.86$\pm 0.39$[p]&0.45\\
NGC 3379 &1.28[b]&4.6&8.63&9.29&8.60$^{+0.10}_{-0.13}$[d]&4.90\\
NGC 3608 &1.29[b]&4.8&7.82&8.50&8.30$^{+0.19}_{-0.16}$[d]&1.57\\
NGC 3640&1.22[a]&4.0&6.35&6.96&7.99$\pm 0.39$[p]&0.09\\
NGC 3706 &1.34[a]&5.2&8.30&9.01&8.94$\pm0.39$[p]&1.18\\
NGC 3842&1.38[a]&5.6&9.20&9.95&9.98$^{+0.12}_{-0.14}$[d]&0.93\\
NGC 4073 &1.17[a]&3.2&8.50&9.00&8.98$\pm0.39$[p]&1.05\\
NGC 4278 &1.26[b]&4.5&8.31&8.96&8.62$\pm0.39$[p]&2.17\\
NGC 4291 &1.27[a]&4.5&8.15&8.81&8.52$^{+0.11}_{-0.62}$[d]&1.94\\
NGC 4365 &1.33[b]&5.0&8.68&9.38&8.81$\pm0.39$[p]&3.70\\
NGC 4382 &1.10[b]&2.6&7.68&8.09&7.95$\pm0.39$[p]&1.37 \\
NGC 4406 &1.25[b]&4.5&8.28&8.93&8.61$\pm0.39$[p]&2.12\\
NGC 4472&1.33[b]&5.0&8.59&9.29&9.15$\pm0.39$[p]&1.40\\
NGC 4552 &1.29[b]&4.8&8.02&8.70&8.67$^{+0.04}_{-0.05}$[d]&1.08\\
NGC 4589 &1.33[b]&5.0&7.50&8.22&8.49$\pm0.39$[p]&0.54\\
NGC 4649&1.34[a]&5.2&9.05&9.76&9.67$^{+0.08}_{-0.10}$[d]&1.22\\
NGC 5061 &1.24[a]&4.2&8.64&9.27&8.05$\pm 0.39$[p]&16.7\\
NGC 5419 &1.35[a]&4.7&9.70&10.37&9.57$\pm0.40$[p]&6.26\\
NGC 5557 &1.18[a]&3.2&8.23&8.74&8.78$\pm0.39$[p]&0.90\\
NGC 5813&1.31[b]&5.0&8.24&8.93&8.83$^{+0.04}_{-0.05}$[d]&1.27\\
NGC 5982 &1.26[b]&4.5&8.24&8.89&8.65$\pm 0.39$[p]&1.75\\
NGC 6849&1.03[a]&2.4&7.97&8.35&8.04$\pm 0.39$[p]&2.03\\
NGC 6876&1.26[a]&4.5&8.52&9.17&8.55$\pm0.39$[p]&4.25\\
NGC 7619&1.35[b]&4.7&8.99&9.66&9.37$\pm0.39$[p]&1.95\\
NGC 7785&1.33[a]&5.0&6.67&7.37&8.80$\pm0.39$[p]&0.04\\
\hline
\end{tabular} 

Notes.---Col. (1) Galaxy name. Col. (2) Galaxy colour: we use the Lauer et al.\ (2005) central $V-I$ colours [b] when available; otherwise the $V-I$ colours [a] were taken from the HyperLeda database. Col. (3) $V$-band stellar mass-to-light ($M/L$) ratios determined using the galaxy colours (col. 2) and the colour-age-metallicity-$(M/L)$ relation given by Graham \& Spitler (2009, their Fig. A1). Col. (4) Central luminosity deficit in terms of $V$-band solar luminosity. Col. (5) Central stellar mass deficit determined using col. (3) and col. (4). Col. (6) SMBH mass. Sources: [p] supermassive black hole mass predicted using the Graham \& Scott (2013, their Table 3 and Fig.~2) ``non-barred $M-\sigma$'' relation (and the ``barred $M-\sigma$'' relation for NGC 6849); [d] galaxies with dynamically determined SMBH mass measurements taken from Graham \& Scott (2013). For NGC 1399, we use two direct (dynamically determined) SMBH mass measurements taken from Gebhardt et al.\ (2007, $M_{\rm BH}=$4.7$^{+0.6}_{-0.6}$$\times$$10^{8}$ $M_{\sun}$) and Houghton et al.\ (2006, $M_{\rm BH}=$1.2$^{+0.58}_{-0.58}$$\times$$10^{9}$$M_{\sun}$) and adjusted to our distance. We use Eq.~4 from Graham et al.\ (2011), updated according to the relation in Graham \& Scott (2013), as well as the $\sigma$ values in Table \ref{Tabbb1} and assume a 10\% uncertainty on $\sigma$ to estimate the error on the predicted SMBH mass (see Graham et al.\ 2011, their Section 2.1.1). Col. (7) Ratio between mass deficit and black hole mass. 
\end {minipage}
\end{table*}
\end{center}

\subsection{Central stellar mass deficits}\label{Sec5.2}

As mentioned before, the central stellar mass deficits of
core-S\'ersic galaxies are naturally generated through the
gravitational sling-shot ejection of core stars by the inspiraling
black hole binaries that that are formed in a merger remnant (Begelman
et al.\ 1980; Ebisuzaki et al.\ 1991). A key point to note is that
high-accuracy simulations (e.g., Milosavljevi\'c \& Merritt 2001;
Merritt 2006) predicted that multiple dissipationless mergers will
have cumulative effects on core formation. Merritt (2006) found that
the total stellar mass deficit, $M_{\rm def}$, after $N$ successive
dry major mergers is $\approx 0.5NM_{\rm BH}$, with $M_{\rm BH}$ the
final SMBH mass.

Past studies have quantified this stellar mass deficit from the
difference in luminosity, $L_{\rm def}$, between the inward
extrapolation of the outer S\'ersic profile (of the core-S\'ersic
model) and a sharp-transition\footnote{The 5-parameter
  sharp-transition core-S\'ersic model is obtained by setting the
  transition parameter $\alpha \rightarrow \infty$ (Graham et
  al.\ 2003).}  core-S\'ersic model (Graham 2004; Ferrarese et
al.\ 2006; Hyde et al.\ 2008). Here, we apply the same prescription
for $L_{\rm def}$ as in these past works but we use a smoother
transition (instead of a sharp) core-S\'ersic model by applying a
finite $\alpha$ in Eq.~\ref{Eqq10} (cf.\ also Dullo \& Graham 2012,
2013). Therefore, the difference in luminosity between the outer
S\'ersic model (Eq.~\ref{Eqq9}) and the core-S\'ersic model
(Eq.~\ref{Eqq37}) is the central stellar luminosity deficit $L_{\rm
  def}$. For each galaxy this luminosity deficit is converted into a
mass deficit using the $V$-band stellar mass-to-light ($M/L$) ratio
given in Table~\ref{Tabbb3}. In order to determine these $M/L $
ratios, the central (if available, otherwise the galaxy) $V-I$ colours
(Table~\ref{Tabbb3}) together with the colour-age-metallicity-($M/L$)
diagram (Graham \& Spitler 2009, their Fig. A1) were used, assuming a
12 Gyr old stellar population. We note that the Graham \& Spitler
(2009) colour-age-metallicity-($M/L$) diagram is constructed using the
Bruzual \& Charlot (2003) stellar population models and the Chabrier
(2003) stellar initial mass function (IMF). Recent works suggest that
the IMF may vary with velocity dispersion for early-type galaxies but
there is a significant scatter in this relation (e.g., Cappellari et
al.\ 2012, 2013; Conroy \& van Dokkum 2012; Spiniello et al.\ 2012;
Wegner et al.\ 2012; Ferreras et al.\ 2013; Zaritsky et
al.\ 2014). Galaxies with low-velocity dispersions require a Kroupa
(2001, or a Chabrier 2003) IMF, while high-velocity dispersion
galaxies ($\sigma \ga 200$ km s$^{-1}$) may prefer a ``bottom-heavy''
IMF having a steeper slope than that of Salpeter (1955), although
Cappellari et al.\ (2013, their Fig.\ 15) found a shallow
($M/L)-$$\sigma$ relation for their slow rotators with $\sigma \ga
160$ km s$^{-1}$ (see also Rusli et al.\ 2013; Clauwens, Schaye \&
Franx 2014). Once these mass-to-light issues are settled, it may be
worth trying to refine the stellar mass deficits reported here.

\begin{figure}
\includegraphics[angle=270,scale=0.50]{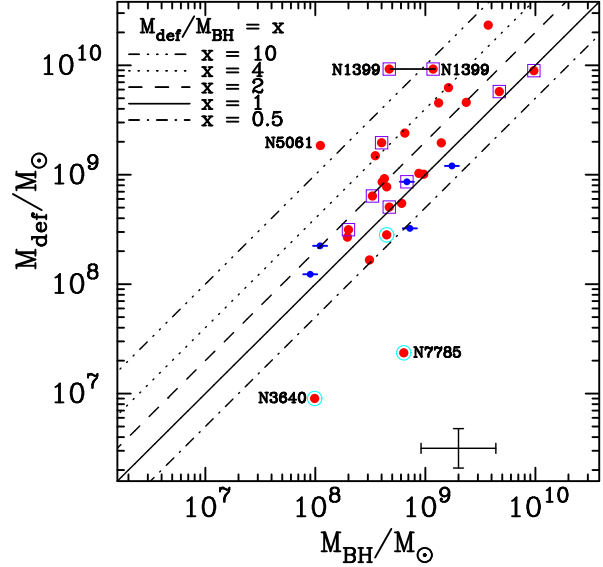}
\caption{Central mass deficit $(M_{\rm def})$ versus black hole mass
  ($M_{\rm BH}$) for the 31 core-S\'ersic galaxies listed in
  Table~\ref{Tabbb1}. The data for four S0s (NGC 507, NGC 2300, NGC
  4382 and NGC 6849) and one elliptical (NGC 3706) are taken from
  Dullo \& Graham (2013). The Graham \& Scott (2013, their Table 3)
  ``non-barred $M$-$\sigma$'' relation was used for estimating
  the SMBH masses of 23 galaxies, while for the
  remaining 8 galaxies (enclosed in boxes) we used
  their direct SMBH mass measurements as given in Graham \& Scott
  (2013). For NGC 1399, we also plot the larger black hole mass from
  Houghton et al.\ (2006) adjusted for our distance of 19.4 Mpc. Three
  questionable galaxies with unusually small cores (NGC 1700, NGC 3640
  and NGC 7785) are circled. A representative error bar is shown at
  the bottom of the panel.}
\label{FigIII12} 
\end{figure}

Fig.~\ref{FigIII12} plots the mass deficits that we derive against the
dynamically determined or predicted SMBH masses for our 31
core-S\'ersic early-type galaxies (Table~\ref{Tabbb3}), the largest
sample of core-S\'ersic galaxies with extended light profiles that has
been modelled to date. We find the mass deficits for these galaxies
are typically $M_{\rm def} \sim 0.5-4$ $M_{\rm BH}$, in agreement with
past core-S\'ersic model estimates (Graham 2004; Ferrarese et al.\
2006; Hyde et al.\ 2008; Dullo \& Graham 2012, 2013). This translates
to core-S\'ersic galaxy formation through one to several successive
``dry'' major merger events, consistent with theoretical expectations
(e.g., Haehnelt \& Kauffmann 2002). In addition, recent observations
on close major merger pairs have revealed that massive galaxies have
undergone $0.5-2$ major mergers since $z\sim 1$ (e.g., Bell et al.\
2004, 2006; Bluck et al.\ 2012; Man et al.\ 2012; Xu et al.\ 2012).
The most massive galaxies, with stellar mass $M_{\star} > 10^{11.5}
M_{\sun}$, may have experienced up to six major mergers since $z \sim
3$ (Conselice 2007).

%As shown in Fig.~\ref{FigIII12b}(a), excluding the three questionable
%galaxies and using the Hougton et al.\ (2006) SMBH mass for NGC 1399,
%the OLS bisector fits the relation between $M_{\rm def}$ and $M_{\rm
%  BH}$ such that
%\begin{equation}
%\mbox{log}\left(\frac{M_{\rm def}}{M_{\sun}}\right)= (1.18\pm 0.16)~\mbox{log}\left(\frac{M_{\rm BH}}{10^{9} M_{\sun}}\right) +~(9.34~ \pm 0.11).~~
%\label{Eqq35def}
%\end{equation}
%This near-linear relation (Eq.~\ref{Eqq35def}) has a vertical scatter
%of 0.37 dex.
% The slope of this relation is slightly greater than 1.
%This is because the $M_{\rm def}/M_{\rm BH}$ ratio is generally larger
%in galaxies with bigger black holes, i.e., in galaxies with bigger
%spheroid masses (Fig.~\ref{FigIII12b}). 

%We however do not find a correlation between $N$ and $M_{\rm BH}$.
% Hever,
%since the extent of core depletion in core galaxies is an indicative
%of the amount of galactic merging as well as the mass of the SMBH in
%the galaxy, it implies a scatter about the linear $M_{\rm def}-M_{\rm
%  BH}$ relation (Fig.~\ref{FigIII12}).

Four elliptical galaxies NGC 1399, NGC 3640, NGC 5061 and NGC 7785 are
outliers from the main $M_{\rm def}-M_{\rm BH}$ distribution
(Fig.~\ref{FigIII12}).  NGC 3640 and NGC 7785 are two of the three
galaxies with unusually small depleted cores (see Section 4 and
Appendix~\ref{ApppQes}).  They both have small mass deficits ($M_{\rm
  def,N3640}\sim 9.0 \times 10^{6} M_{\sun}$ and $ M_{\rm def,N7785}
\sim 2.4\times10^{7} M_{\sun}$) for their predicted SMBH masses
($M_{\rm BH,N3640}\sim 9.8\times10^{7} M_{\sun}$ and $M_{\rm
  BH,N7785}\sim 6.4\times10^{8} M_{\sun}$). The remaining galaxy with
a questionable core (NGC 1700) is an outlier in most central galaxy
scaling relations (Figs.~\ref{FigIII6}, \ref{FigIII7} and
\ref{FigIII8}) but it has a normal mass deficit for its SMBH mass.
This owes to the fact that NGC 1700, unlike NGC 3640 and NGC 7785, has
a relatively steep outer S\'ersic profile ($n=6.1$). This larger
S\'ersic index value helps to compensate for the small core size,
taking its estimated mass deficit into the normal range in
Fig.~\ref{FigIII12}.

In the case of the potentially outlying galaxy NGC 1399, the
discussion given in Section~\ref{Sec5.1} explains the behavior seen
here. The smaller dynamical SMBH mass determination by Gebhardt et
al.\ (2007) yields an inflated $M_{\rm def}/M_{\rm BH}$ ratio of $
\sim 20$, while assuming the larger dynamical SMBH mass measurement of
Houghton et al.\ (2006) implies a somewhat reasonable value of $M_{\rm
  def}/M_{\rm BH} \approx 8$.

The situation with the fourth offset elliptical galaxy NGC 5061 is
somewhat unclear given it is not a deviant galaxy in the other galaxy
scaling relations (Section~\ref{Sec4}). The core-S\'ersic model fits
its light profile very well with a fairly small rms residual of $\sim
0.05$ mag arcsec$^{-2}$, but from these fit parameters we determine a
large mass ratio $M_{\rm def}/M_{\rm BH} \approx 17$. Its core size
$R_{b}=34$ pc is a good match to its $V$-band absolute magnitude
$M_{V}\sim -22.5$ mag (Table~\ref{Tabbb1}) and velocity dispersion
$\sigma =186$ km s$^{-1}$ (Table~\ref{Tabbb1}), but its S\'ersic index
$n\sim 8.4$ (Table~\ref{Tabbb2}) may be too high for the
aforementioned galaxy properties. Indeed, NGC 5061 has the largest
S\'ersic index from our sample, attributed to its noticeably straight
surface brightness profile (Fig.~\ref{FigA1}). This may suggest that
the envelope of this galaxy was built via several dry minor and major
merging events (Hilz et al.\ 2013). On the other hand, it has the
third smallest velocity dispersion $\sigma=186$ km s$^{-1}$
(HyperLeda's mean value) from our sample which seems to underpredict
its SMBH mass $M_{\rm BH}=1.1 \times 10^{8} M_{\sun}$
(Table~\ref{Tabbb3}). Using the largest reported velocity dispersion
value $\sigma=213$ km s$^{-1}$ (Davies et al.\ 1987), instead of the
mean measurement, increases its predicted SMBH mass roughly by a
factor of 2, i.e., $M_{\rm BH}=2.4 \times 10^{8} M_{\sun}$. As such
the associated $M_{\rm def}/M_{\rm BH}$ ratio reduces roughly by a
factor of two, to give $M_{\rm def}/M_{\rm BH} \sim 7.9$. This latter
ratio is marginally consistent with the $M_{\rm def}/M_{\rm BH}$
distribution shown in Fig.~\ref{FigIII11}. In summary, it appears that
both the high S\'ersic index and the relatively low velocity
dispersion of NGC 5061 {\it may} collectively act to inflate the
$M_{\rm def}/M_{\rm BH}$ ratio to 17.

\begin{figure}
\includegraphics[angle=270,scale=0.82]{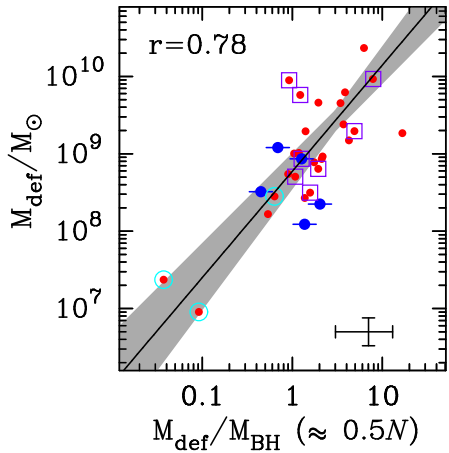}
\caption{Central mass deficit $(M_{\rm def})$ versus cumulative number
  of ``dry'' major merger events ($N$) for the 31 core-S\'ersic
  galaxies listed in Table~\ref{Tabbb1}. Symbolic representations are
  as in Fig.~\ref{FigIII12}. Pearson correlation coefficient, $r$,
  (and representative error bar) are shown at the top (bottom) of the
  panel.}
\label{FigIII12b} 
\end{figure}

\begin{figure}
\includegraphics[angle=270,scale=0.480]{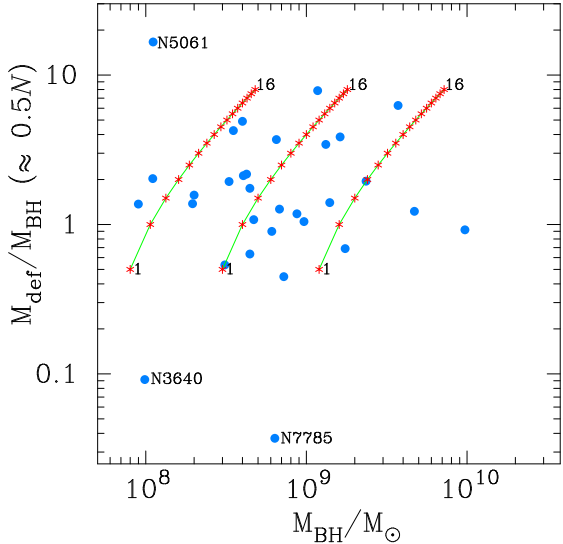}
\caption{The ratio of central stellar mass deficit $(M_{\rm def})$ to
  supermassive black hole mass ($M_{\rm BH}$) as a function of $M_{\rm
    BH}$ for the 31 core-S\'ersic galaxies listed in Table~\ref{Tabbb1}.
  The three curves are based on the simulations by Merritt (2006, his
  Table 2). That is, we started with a certain (total) supermassive
  black hole mass $M_{\rm BH}$ for $N=1$ and a third of this black
  hole mass was added for each successive merger shown by a star. As
  such, the cumulative black hole mass $M_{\rm BH}$ increases linearly
  with the number of merger $N$, but the binary mass ratio decreases
  with $N$. These black hole masses were then plotted against
  $0.5N$($\approx$ $M_{\rm def}/M_{\rm BH}$, Merritt 2006) to
  construct each curve, see the text for further detail.}
\label{FigIII12c} 
\end{figure}

As shown in Fig.~\ref{FigIII12b}, excluding the three galaxies with
questionably small/real cores, and using the Hougton et al.\ (2006)
SMBH mass for NGC 1399, the OLS bisector regression between $M_{\rm
  def}$ and $M_{\rm def}/M_{\rm BH}$ ($\approx$ $0.5N$, Merritt 2006)
gives
\begin{equation}
\mbox{log}\left(\frac{M_{\rm def}}{M_{\sun}}\right) = (1.37\pm 0.20)~\mbox{log}\left(\frac{0.5N}{3.0}\right) +~(9.43~ \pm 0.12).~~
\label{Eqq35def2}
\end{equation}
In Fig.~\ref{FigIII12c}, we explore the behavior of the $M_{\rm
  def}/M_{\rm BH}$ ratio with $M_{\rm BH}$. We find that the
distribution in this diagram appears largely consistent with the
simulations by Merritt (2006, his Table 2). As in Merritt (2006), for
the first merger ($N=1$), an object with a supermassive black hole
mass of (1/3)($M_{BH,N=1}$) was added to a system having a black hole
mass of (2/3)($M_{BH,N=1}$). The same black hole mass of
(1/3)($M_{BH,N=1}$) was then added for each successive merger. These
accumulated black hole masses are plotted against $0.5N$ to construct
each of the three curves shown in Fig.~\ref{FigIII12c}. Form this
figure it is apparent that core-S\'ersic galaxies with the same SMBH
mass (or merger history) can have different merger histories (SMBH
masses). This is consistent with the notion that the stellar mass
deficits of core-S\'ersic galaxies reflect the amount of merging as
well as the masses of their SMBHs.

\subsection{Comparison with past stellar mass deficit measurements}\label{Sec5.3}
%As noted in passing, our result $M_{\rm def}/M_{\rm BH} \sim0.5-4$ is
%in an excellent agreement with those from previously published
%core-S\'ersic works (Graham 2004; Ferrarese et al.\ 2006; Hyde et al.\
%2008; Dullo \& Graham 2012, 2013).  
The mean elliptical galaxy
$M_{\rm def}/M_{\rm BH}$ ratio from Graham (2004) is 2.1 $\pm$ 1.1,
while Ferrarese et al.\ (2006) reported a mean $M_{\rm def}/M_{\rm
  BH}$ ratio of 2.4 $\pm$ 0.8 after excluding the S0 galaxy NGC 4382
from their sample.  Hyde et al.\ (2008) found a comparable mean
$M_{\rm def}/M_{\rm BH}$ ratio of 2.3 $\pm$ 0.67 for their sample. In
Dullo \& Graham (2012), we modelled $\sim$$10\arcsec$ light profiles
and cautioned that the outer S\'ersic parameters might be less
constrained than desirable, although the S\'ersic indices were shown
to be in a fair agreement with those determined from published fits to
larger radial extents. Nonetheless, we reported tentative $M_{\rm
  def}/M_{\rm BH}$ ratios that were some 0.5$- 4$. In Dullo \& Graham
(2013) we fit the extended light profiles of four core-S\'ersic
lenticular galaxies (NGC 507; NGC 2300; 4382 and NGC 6849) using a
core-S\'ersic model for the bulge plus an exponential model for the
disc. One suspected S0 galaxy NGC 3706 was found to have a stellar
distribution that is best described by the core-S\'ersic model and was
thus reclassified as an elliptical galaxy. Using these core-S\'ersic
fit parameters we reported a robust $M_{\rm def}/M_{\rm BH}=0.5 -2$
for these five core-S\'ersic galaxies (Dullo \& Graham 2013, their
Fig.~4), which are also shown here in Fig.~\ref{FigIII12}.

%The
%existence of stellar mass deficits in core-S\'ersic lenticular
%galaxies suggests a two stage inside-out mechanism for their assembly.
%First, a violent ``dry'' merger event creates a bulge with a depleted
%core before a disc gradually grows around the naked bulge via cold gas
%accretion (e.g., Steinmetz \& Navarro 2002; Kere\v{s} et al.\ 2005;
%Bouch\'e et al.\ 2013).

Prior to Graham (2004) who reported $M_{\rm def}/M_{\rm BH}$ ratio of
$\sim 2$, previous estimates based on Nuker model parameters had been
an order of magnitudes larger (e.g., Milosavljevic \& Merritt 2001;
Milosavljevi\'c et al.\ 2002; Ravindranath 2002). Graham (2004) argued
that the Universe was some ten times less violent, in terms of major
galaxy mergers, than previously believed. Subsequent works using Nuker
model parameters (Lauer et al.\ 2007a; G\"ultekin et al.\ 2011) or
subjectively identifying the core from visual inspections (Kormendy et
al.\ 2009) have reported mass deficits up to an order of magnitude
larger than typically found here.  As discussed in Section \ref{Sec3}
and Appendix~\ref{ApppC}, this discrepancy is partly due to the
contrasting core sizes measured by these distinct methodologies. Using
a model-independent analysis of the light profiles, Hopkins \&
Hernquist (2010) confirmed the result of Graham (2004) and reported
$M_{\rm def}/M_{\rm BH} \approx 2$. The larger mass deficits ($M_{\rm
  def} \sim 5-20~M_{\rm BH}$) of Kormendy \& Bender (2009) are also
partly because they used the dynamical mass-to-light $(M/L)_{\rm dyn}$
ratios rather than the stellar $M/L$ ratios to derive the mass
deficits.  There is not much dark matter at the centre of massive
spheroids (e.g., Dekel et al.\ 2005), and assuming the dynamical
$(M/L)_{\rm dyn}$ for the galaxy (= [$M_{\rm stars}$ + $M_{\rm
  dark~matter}]$ / $L$) overpredicts the central mass deficits.
Recently, Kormendy \& Ho (2013) scaled up the SMBH masses for the
galaxies in Kormendy \& Bender (2009) by about a factor of two and
reported a new lower, mean ratio $M_{\rm def}/M_{\rm BH}= 4.1$.

Alternative mechanisms for the production of enhanced depleted cores
in luminous galaxies has been suggested in the literature
(Boylan-Kolchin et al.\ 2004; Gualandris \& Merritt 2008; Kulkarni \&
Loeb 2012).  Gualandris \& Merritt (2008) invoked the recurrent core
passages of gravitationally kicked and ``recoiled'' (and then
oscillating about the centre) SMBH in a merger remnant to explain the
formation of large stellar mass deficits that are up to 5 $M_{\rm
  BH}$. Kulkarni \& Loeb (2012) also suggested that mass deficits as
large as $5 M_{\rm BH}$ could be formed as a result of the action of
multiple SMBHs from merging galaxies. If these processes have always
occurred, then the result we found here, i.e., $M_{\rm def} \sim
0.5-4$ $M_{\rm BH}$ for our sample galaxies implies that these
galaxies are formed via just one major merger or minor merger events
only, at odds with both observations of close galaxy pairs of equal
mass, and theoretical expectations (e.g., Khochfar \& Burkert 2003;
Bell et al.\ 2004; Naab et al.\ 2006; Bluck et al.\ 2012). However,
spheroids with $N > 8$~(i.e., $ M_{\rm def}/M_{\rm BH} > 4)$ probably
need some black hole oscillations as 8 or more major mergers seem
excessive.  We find two elliptical galaxies (NGC 1399, NGC 5061) with
$M_{\rm def}/M_{\rm BH} \ga 8$, suggesting that these oversized mass
deficits might be partly due to the action of their gravitationally
kicked SMBHs.
% observations have revealed that compact ($R_{e} \la 2$ kpc)
%massive high red shift ($z \sim 2$) galaxies, which are often thought
%to be the plausible precursors of today's massive ellipticals, are up
%to a factor of 5 smaller than local elliptical galaxies of comparable
%stellar mass (e.g., Daddi et al.\ 2005; Trujillo et al.\ 2006).
%However, given the shortage of satellites, both dissipationless minor
%and major merging events are necessary to explain this large size
%expansion over the last 10 billion years (e.g., Ciamatti et al.\ 2012;
%Trujillo 2012).
%Also, the near-linear relations between BH mass and host spheroid
%mass, and luminosity for the core-S\'ersic galaxies (Graham 2012;
%Graham \& Scott 2013) suggest galaxy formation via the self-addition
%of comparable mass (gas-free) systems, not minor mergers.

\subsection{Alternative core formation scenarios}\label{Sec5.4}
It should be noted that several authors have considered
alternative ways in which cores can be produced in luminous galaxies.
Dissipationless collapses in existing dark matter haloes were invoked
by Nipoti et al.\ (2006) as a possible mechanism for forming depleted
cores. Another suggested alternative was the adiabatic expansion of
the core region driven by the rapid mass loss from the effects of
supernova and AGN feedback (Navarro et al.\ 1996; Read \& Gilmore
2005; Peirani, Kay \& Silk 2008; Martizzi et al.\ 2012, 2013) and
Krajnovi\'c et al.\ (2013) pointed out that this scenario is
compatible with the properties of ``core slow rotators''. In addition,
Goerdt et al.\ (2010) proposed that the energy transferred from
sinking massive objects would produce cores that are as large as 3 kpc
in size. These suggested mechanisms, however, are not without
problems. For example, it is unclear how the cores created by the
simulations in Nipoti et al.\ (2006) are guarded against infalling
satellites (which would replenish the core) in the absence of a
central SMBH. Also, the oversized ($3-10$ kpc) cores produced by the
latter two mechanisms (e.g., Goerdt et al.\ 2010; Martizzi et al.\
2012, 2013) are generally inconsistent with the typical $\la$0.5 kpc
cores observed in galaxies (e.g., Trujillo et al.\ 2004; Ferrarese et
al.\ 2006; Richings et al.\ 2011; Dullo \& Graham 2012). Section 6.1
of Dullo \& Graham (2013) provides further details, including the
merits and weaknesses of these core formation models in the context of
the observations.

\subsection{Comparison of core-S\'ersic elliptical and lenticular galaxies}

Finally, we note that the standard Lambda Cold Dark Matter
($\Lambda$CDM) model predicts that elliptical galaxies are built via
major mergers (e.g., Kauffmann et al.\ 1993; Khochfar \& Burkert
2005). In this hierarchical picture, the bulges of lenticular galaxies
form early via major mergers while their discs grow later through gas
accretion events (e.g., Steinmetz \& Navarro 2002). Alternatively, the
evolutionary transformation of spiral galaxies into S0 galaxies via
mechanisms such as ram pressure stripping has been suggested (e.g.,
Gunn \& Gott 1972). Comparing various galaxy scaling relations,
Luarikainen et al.\ (2010), for example, showed that the bulges of S0 galaxies
are closely correlated with the bulges of bright spiral galaxies, having
$M_{K}$(bulge) $< -$20 mag, than with elliptical galaxies. On the
other hand, in Dullo \& Graham (2013), we argued that core-S\'ersic
S0s with $M_{V}$(bulge) $\la -21.30$ mag might be assembled inside-out
in two stages: an earlier ``dry'' major merger process involving SMBHs
forms their bulge component, while the surrounding disc is
subsequently formed via cold gas accretion. The bulges of our small S0
galaxy sample tend to have $n \la 3$ (Dullo \& Graham 2013, see also
Balcells et al.\ 2003 and Laurikainen et al.\ 2005), however,
core-S\'ersic elliptical galaxies have $n\ga4$. As we mentioned in
Section~\ref{Sec4.1}, for the same core size, these massive bulges
tend to be compact ($R_{e} \la 2$ kpc, Fig.~\ref{FigIII7}b) and have
somewhat fainter break, and central, surface brightnesses than the
elliptical galaxies (Figs.~\ref{FigIII6}a and
\ref{FigIII6}b). Furthermore, the $M_{\rm def}/M_{\rm BH}$ ratio for
S0s is lower compared to the spread seen in elliptical galaxies
(Fig.~\ref{FigIII12}). This implies that the bulges of core-S\'ersic
S0s have experienced fewer major mergers than core-S\'ersic elliptical
galaxies (Fig.~\ref{FigIII12b}).

%\begin{figure}
%\includegraphics[angle=270,scale=0.45]{MVdef_vs_BHIII.ps}
%\caption{Absolute magnitude of the stellar deficit}
%\label{FigIII4} 
%\end{figure}

\section{Conclusions}\label{Sec6}

We extracted the major-axis surface brightness profiles from 26
core-S\'ersic early-type galaxies observed with the {\it HST} WFPC2
and ACS cameras. We additionally included five core-S\'ersic
early-type galaxies (NGC 507, NGC 2300, NGC 3706, NGC 4382 and NGC
6849) from Dullo \& Graham (2013). This compilation represents the
largest number of core-S\'ersic galaxies modelled to large radii $R
\ga 80\arcsec$, giving the fitting functions enough radial expanse to
robustly measure the galaxy stellar distribution (see
Fig.~\ref{FigIII5} for a comparison of the S\'ersic indices obtained
from fits using $10-15 \arcsec$ profiles).  We fit the extended surface
brightness profiles of the 26 core-S\'ersic elliptical galaxies using
the core-S\'ersic model, while light profiles of the remaining five
core-S\'ersic S0 galaxies were modelled with the core-S\'ersic model
for the bulge plus an exponential model for the disc. We accounted for
additional nuclear cluster light using the Gaussian function.  Our
principal results are summarised as follows:

1. The global stellar distributions of core-S\'ersic elliptical
galaxies are robustly represented with the core-S\'ersic model, while
core-S\'ersic lenticular galaxies are accurately described using the
core-S\'ersic bulge + exponential disc model. These fits yield a
median rms scatter of 0.045 mag arcsec$^{-2}$ for our sample of 31
core-S\'ersic galaxies, and argue against excessive multi-component
S\'ersic models (Section 3.4).

2. We provide updated core-S\'ersic model parameters $R_{b}, \gamma,
\mu_{b}, n, R_{e}$ for 31 core-S\'ersic early-type galaxies with
spheroidal components having $M_{V} \la -20.7$ mag and $\sigma \ga
179$ km s$^{-1}$. In general, there is a good agreement with the
parameters obtained from our earlier analysis of the publicly
available, but radially limited ($R\la 10\arcsec$) surface brightness
profiles given by Lauer et al.\ (2005).

3. The bulges of our core-S\'ersic S0s are compact ($R_{e} \le $ 2 kpc)
and have $2 < n < 3$ (Dullo \& Graham 2013), as compared to the
core-S\'ersic elliptical galaxies which typically have $5 < R_{e} <
20$ kpc and $n \ga 4$.

4. The core-S\'ersic model break radii are in agreement with both
(i) the previously published core-S\'ersic break radii and (ii) the
model-independent break radii which mark the locations where the
negative logarithmic slopes of the light profiles equal 0.5 (Carollo
et al.\ 1997a; Dullo \& Graham 2012).

5. Updated structural parameter relations involving both the central
and global galaxy properties are provided in Section \ref{Sec4}.
% Using
% the core-S\'ersic parameters, and $(M/L)_{\rm dyn} \propto L^{0}$ for
% core-S\'ersic spheroids rather than $(M/L)_{\rm dyn} \propto L^{1/4}$,
% we have re-derived the Faber et al.\ (1997) central scaling relations
% that are given as a function of spheroid luminosity $L$ and mass
% $M_{\rm dyn}$.
% suggesting the core Fundamental Plane (Faber et al.\
%1997). As noted by Faber et al.\ (1997), the cores of less luminous
%(massive) core-S\'ersic galaxies are much smaller and denser than the
%brighter (more massive) ones.
We have found tight correlations involving the central galaxy
properties $R_{b}$, $\mu_{0}$, $\mu_{b}$, $\sigma$ and $M_{\rm BH}$
(see Table~\ref{Tabbb30}). We have also found near-linear relations
between the break radius $R_{b}$, and the spheroid luminosity $L$ and
the SMBH mass $M_{\rm BH}$ given by $R_{b}\propto L^{1.13 \pm 0.13}$
and $R_{b}\propto M_{\rm BH}^{0.83 \pm 0.21}$.  We additionally found
a near-linear relation between $R_{b}$ and $R_{e}$ such that
$R_{b}\propto R_{e}^{0.98\pm 0.15}$ but with a large scatter.

%(i.e., $R_{b} \approx 0.01R_{e}$, Eq.~\ref{Eqq18}) for the
%core-S\'ersic elliptical galaxies, consistent with the result
%$R_{b}\approx 0.02^{+0.025}_{-0.01} R_{e}$ by C\^ot\'e et al.\ (2007)
%within the errors.

6. We have derived central stellar mass deficits in 31 early-type galaxies
that are typically 0.5 to 4 times the host galaxy's black hole mass.
Given published theoretical results, these mass deficits suggest a few
dissipationless major mergers for core-S\'ersic galaxies.

7. As noted in Dullo \& Graham (2013), mass deficits in core-S\'ersic
S0s suggest a two stage assembly: an earlier ``dry'' major merger
event involving SMBHs creates the bulges with depleted cores, and the
disc subsequently builds up via cold gas accretion events.

%10. The tight correlation between the stellar mass deficit $M_{\rm
%  def}$ and the SMBH mass $M_{\rm BH}$ for core-S\'ersic galaxies is
%such that $M_{\rm def} \propto M_{\rm BH}^{1.18\pm 0.16}$. The slope
%of this near-linear $M_{\rm def}-M_{\rm BH}$ relation is slightly
%bigger than 1, this is because the $M_{\rm def}/M_{\rm BH}$ ratio is
%bigger in galaxies with bigger black holes. 

8. The relation between the stellar mass deficit $M_{\rm def}$ and
the cumulative number  $N$ of major ``dry'' mergers that the galaxy has
undergone is such that $M_{\rm def}$ is roughly $\propto$ $N^{1.37
  \pm 0.20}$.

9. The close relation between the galaxy cores and the SMBHs supports
the popular core depletion hypothesis where cores are thought to be
created by sinking binary SMBHs that eject stars away from the centres
of their host galaxies. The small cores seen in some galaxies, if
real, may arise from loss cone regeneration by newly produced stars
and/or recent stellar accretion events. Alternatively, small galaxy
cores can be interpreted as a sign of minor mergers.

10. We have identified two galaxies (NGC 1399 and NGC 5061) which have
a high $M_{\rm def}/M_{\rm BH}$ ratio, suggesting that their central
SMBH may have experienced a kick due to a gravitational-radiation
recoil event leading to multiple core passages.

\section{Acknowledgments}

This research was supported under the Australian Research Council’s
funding scheme (DP110103509 and FT110100263). This research has made
use of the NASA/IPAC Extragalactic Database (NED) which is operated by
the Jet Propulsion Laboratory, California Institute of Technology,
under contract with the National Aeronautics and Space Administration.
We acknowledge the usage of the HyperLeda database
(http://leda.univ-lyon1.fr.). BTD is grateful for the SUPRA
scholarship offered by Swinburne University of Technology, and travel
support from the Astronomical Society of Australia.

\appendix
\section {}\label{ApppA}
Figure \ref{FigA1} shows core-S\'ersic model fits to the major-axis
surface brightness profiles of the 26 (suspected elliptical)
core-S\'ersic galaxies listed in Table \ref{Tabbb1}. Notes on two of
these 26 galaxies (NGC 4073 and NGC 6876) with complicated structures
are given below, while their photometric profiles are shown in
Figs.~\ref{FigB1} and \ref{FigB2}.

\renewcommand\thefigure{\thesection\arabic{figure}} 
\setcounter{figure}{0} 
\begin{figure*}
\includegraphics[angle=270,scale=0.76]{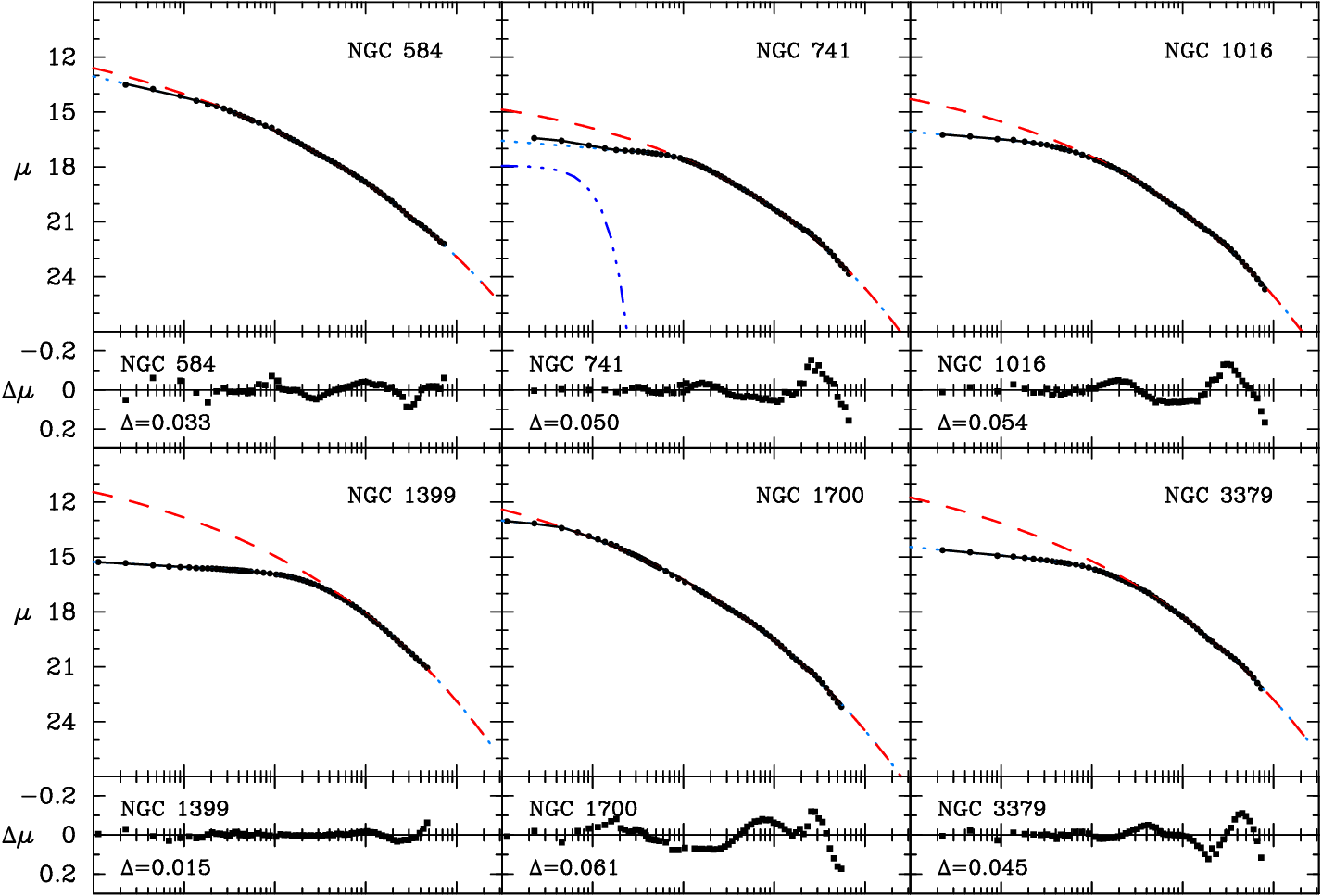}
\includegraphics[angle=270,scale=0.755]{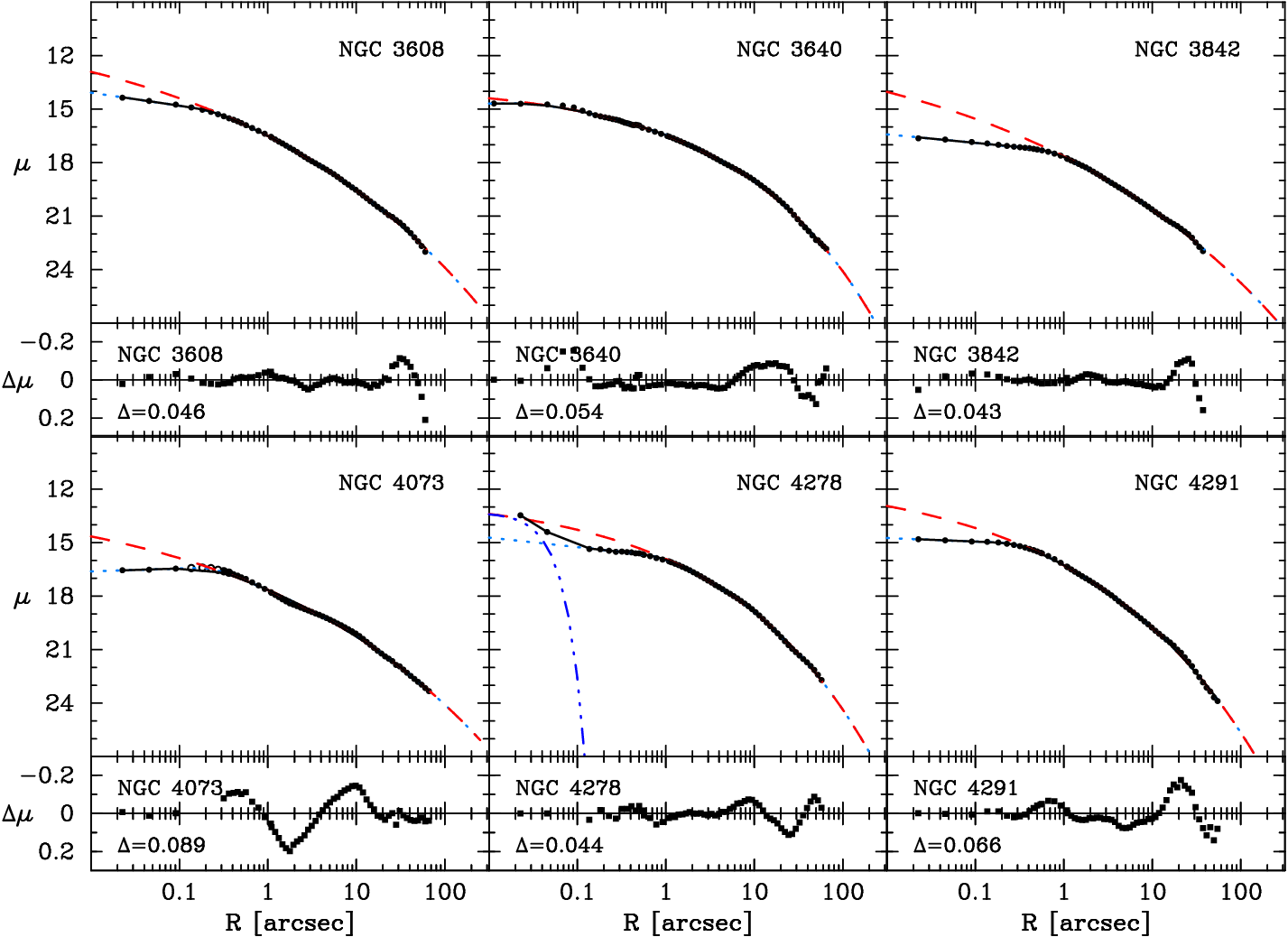}
\caption{
  Fits to the $V$-band, major-axis, surface brightness profiles of the
  elliptical galaxies listed in Table 1, except for NGC 1399 for which
  we fit its $R$-band profile. The long dashed curves indicate the
  S\'ersic component of the fitted core-S\'ersic model (short dashed
  curve). Additional nuclear light components such as AGN and star
  clusters are accounted for with a Gaussian (triple dot-dashed curve)
  function. The solid curves represent the complete fit to the
  profiles, with the rms residuals, $\Delta$, about each fit given in
  the lower panels.}
\label{FigA1}
\end{figure*}
\setcounter{figure}{0}
\begin{figure*}
\includegraphics[angle=270,scale=0.76]{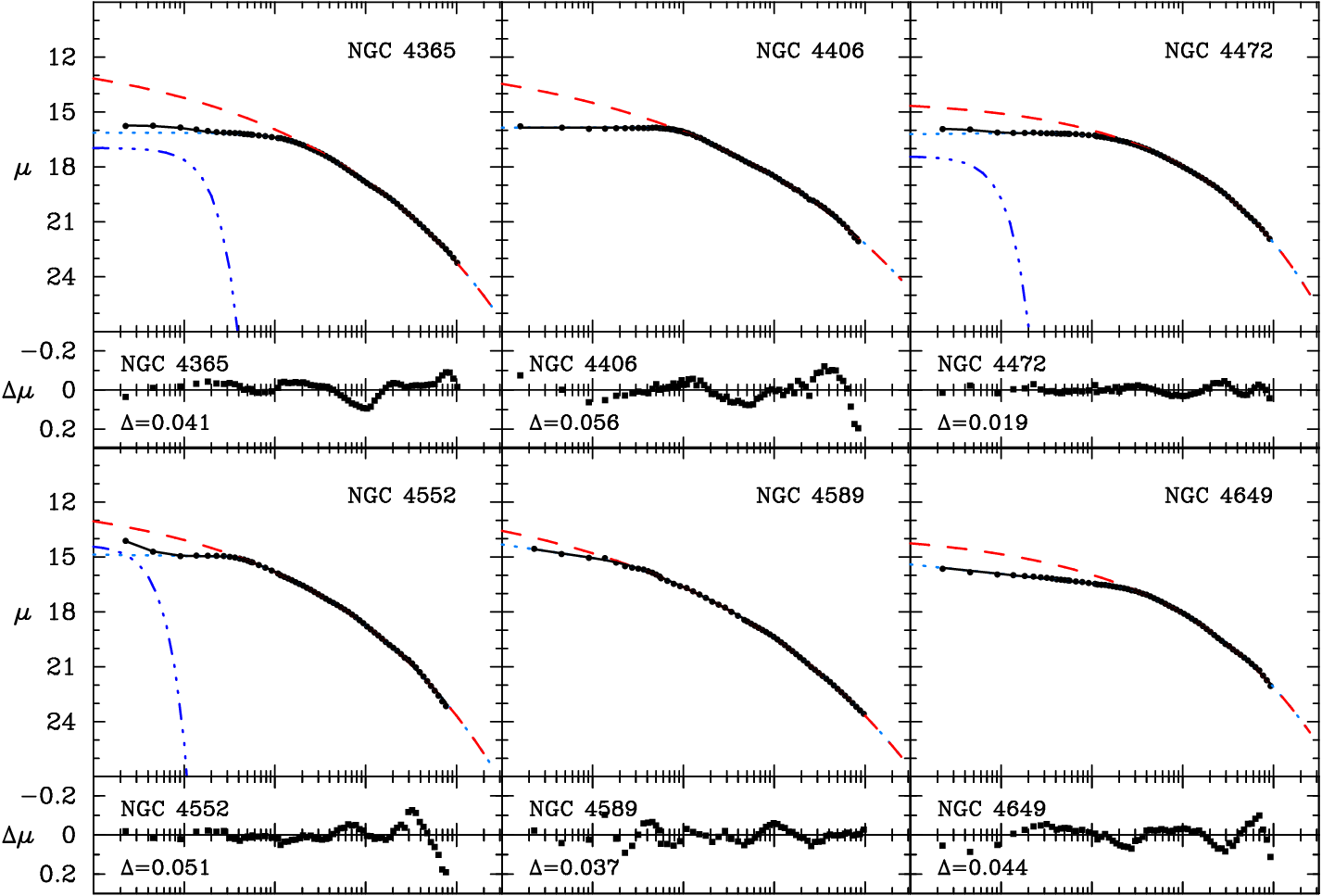}
\includegraphics[angle=270,scale=0.76]{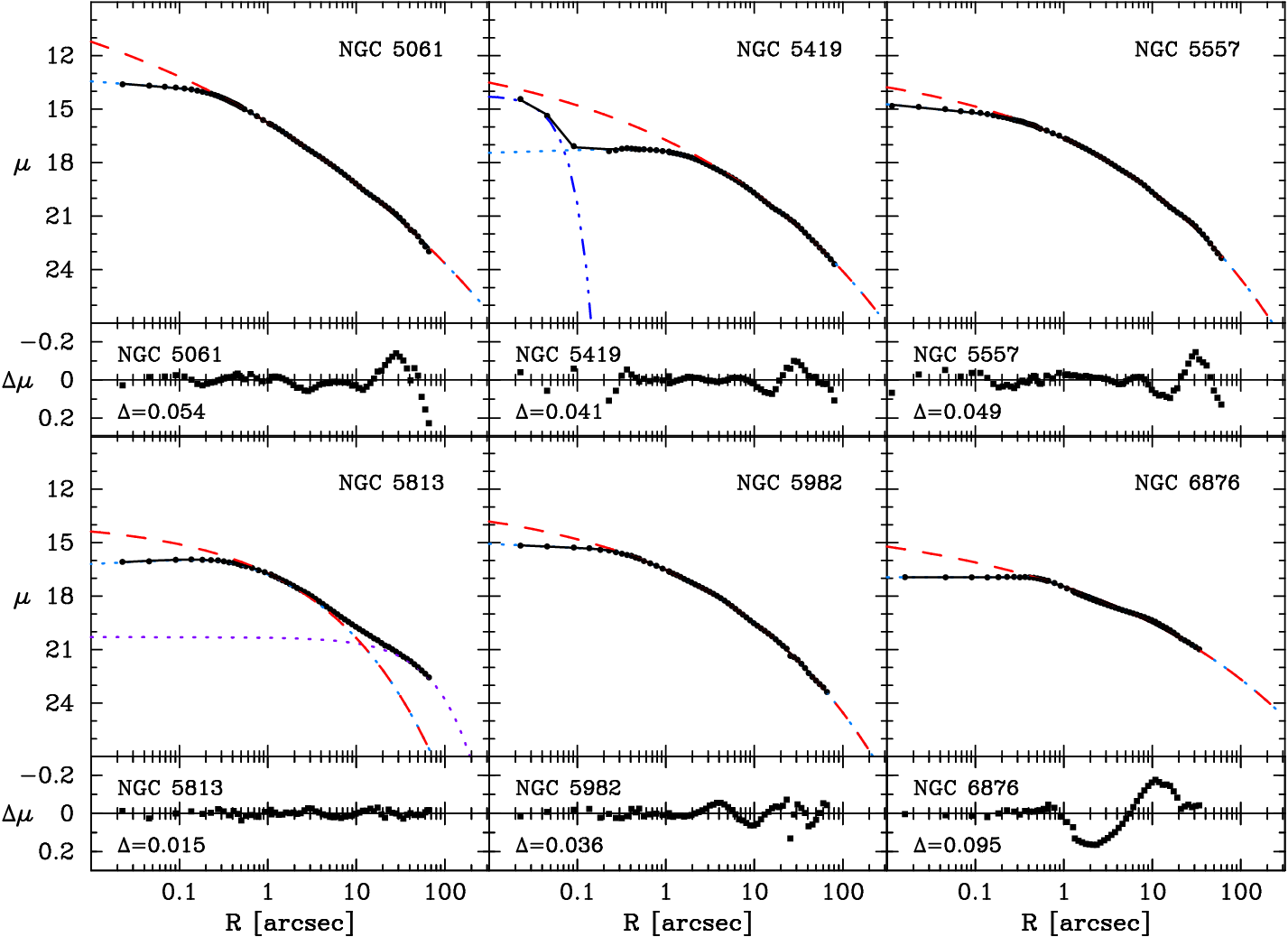}
\caption{(continued)}

\setcounter{figure}{0}
\end{figure*}
\begin{figure*}
\includegraphics[angle=270,scale=0.76]{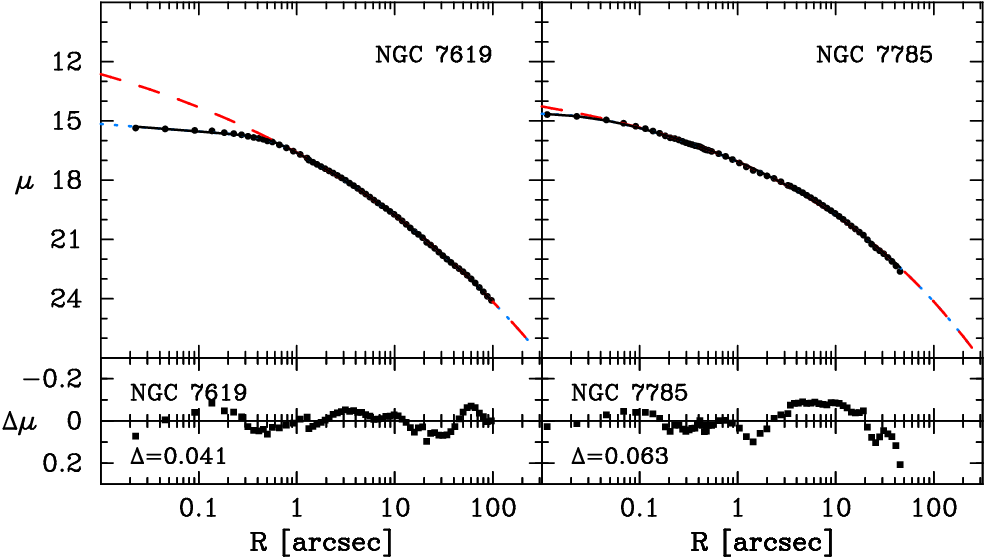}
\caption{(continued)}
\label{Figggkk2} 
\end{figure*}

\subsection{NGC 4073} 

\begin{figure}
\includegraphics[angle=270,scale=0.45]{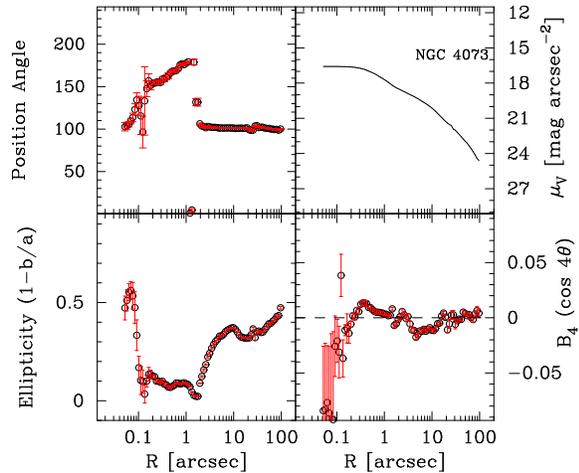}
\caption{Major-axis surface brightness, position angle, ellipticity
  and isophote shape parameter ($B_{4}$) profiles for NGC 4073.}
\label{FigB1}
\end{figure}

NGC 4073 is a cD galaxy in the poor MKW 4 cluster (de Vacouleurs et
al.\ 1991). It has a double classification in Laurikainen et
al.\ (2011, their Table 3), i.e., SAB0$^{-}$ for the inner regions and
E$^{+}5$ for the outer parts. Also, this galaxy has a small bump in
its light profile over $0\arcsec.1<R<0\arcsec.4$ due to a nuclear ring
of stars (Lauer et al.\ 2005). Our models are not designed to fit
stellar rings, thus we simply exclude data points that are
contaminated by the nuclear ring light.  Further, the core-S\'ersic
model fit to NGC 4073 shows an excess of light on top of the
core-S\'ersic light distribution over $2\arcsec < R < 10 \arcsec$,
creating the residual pattern seen in Fig.~\ref{FigA1}. cD galaxies
are known to grow via cannibalism of their neighboring, less massive
cluster galaxies (e.g., Ostriker \& Hausman 1977). If these accreted
objects (or at least their dense cores) survive, they would be visible
as extra light like in NGC 4073.  However, we did not find clear
evidence for such a feature in this galaxy's 2D residual image,
possibly suggesting a collection of large scale disturbances in the
galaxy. Fig.~\ref{FigB1} shows the surface brightness and photometric
profiles for NGC 4073 determined using the IRAF task {\sc
  ellipse}. These profiles are connected with the galaxy's residual
structure observed in the model fit (Fig.~\ref{FigA1}). The position
angle profile shows an abrupt 90$^{\circ}$ twist at around $1
\arcsec$.  The galaxy is also highly flattened ($0.4 < \epsilon < 0.5
$) outside $10 \arcsec$, while the ellipticity shows a steady drop
from 0.45 (at $R \sim 10\arcsec$) to 0.1 (at $R \sim 1\arcsec$)
accompanied by negative B$_{4}$ values, then becomes nearly circular
($\epsilon \sim 0.1$) inside the core ($R < 1\arcsec$). Fisher,
Illingworth \& Franx (1995) reported NGC 4073 as having a
counterrotating stellar core which they attribute to possible
cannibalism events. Laurikainen et al.\ (2011) interpreted the excess
light and the associated ellipticity trend over $2\arcsec \la R \la
10\arcsec$ as being due to a weak inner bar. Without accounting for
its light excess at $1\arcsec<R<10\arcsec$, and omitting its ring
light, the core-S\'ersic model fits the light distribution of NGC 4073
with rms residual of $\sim 0.09$ mag arcsec$^{-2}$.
\subsection{NGC 6876} 

\begin{figure*}
\includegraphics[angle=0,scale=0.30]{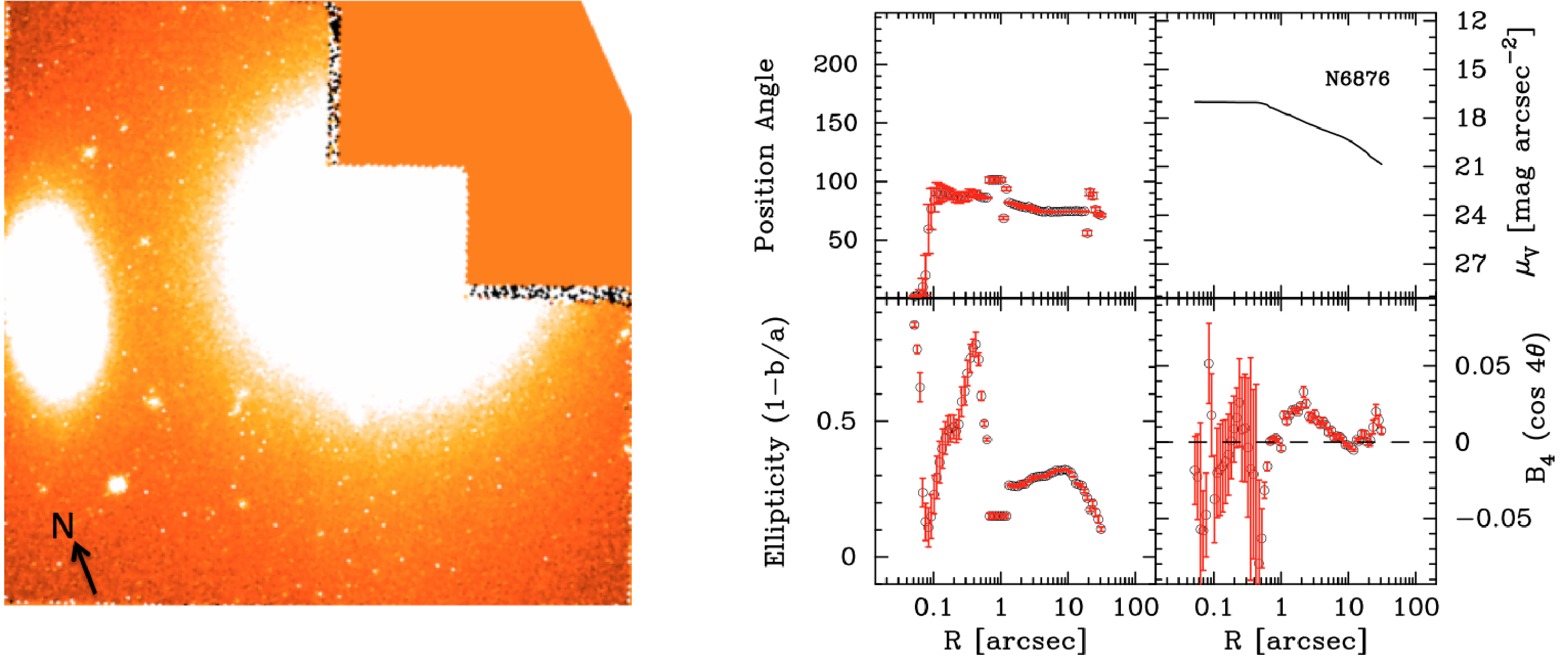}
\caption{Left: {\it HST}/WFPC2 F555W image of the dominant elliptical
  galaxy NGC 6876 centered on the PC1 CCD of the WFPC2 camera. The
  companion elliptical galaxy NGC 6877 sits $1\arcmin.4$ to the east
  of NGC 6876. Right: Major-axis surface brightness, position angle,
  ellipticity and isophote shape parameter ($B_{4}$) profiles for NGC
  6876.}
\label{FigB2}

\end{figure*}

NGC 6876 is a dominant elliptical galaxy in the Pavo group. As shown
in Fig.~\ref{FigB2}, this galaxy is only $1\arcmin.4$ from its smaller
companion, the elliptical galaxy NGC 6877. Using multi-wave
observations, Machacek et al.\ (2005, 2009) showed evidence for
interaction between NGC 6876 and the highly disturbed spiral galaxy
NGC 6872. Dullo \& Graham (2012, their Fig.~4) showed the double
optical nucleus in NGC 6876 possibly associated with the dense core of
a lesser galaxy or the ends of an inclined ring (Lauer et al.\ 2002).
The residual structure at $R \ge1\arcsec$ (Fig.~\ref{FigA1}) is
associated with a 13$^{\circ}$ twist in the position angle, and the
change in the ellipticity and the isophote shape of this galaxy beyond
its core regions.


\section{}\label{ApppQes}

\subsection{Questionable cores}

\noindent {\bf NGC  1700}\label{Sec8.1.1} \\

\noindent NGC 1700, a luminous ($M_{V}=-22.0$ mag) elliptical galaxy
with an estimated (luminosity weighted) age of 3$\pm$1 Gyr (Brown et
al.\ 2000) and a velocity dispersion $\sigma = 239$ km s$^{-1}$, shows
post merger morphological signatures such as shells, and boxy isophote
at large radii $R> 40 \arcsec$ (Franx et al.\ 1989; Statler et
al.\ 1996; Whitmore et al.\ 1997; Brown et al.\ 2000; Stratler \&
McNamara 2002).  It was also reported to be offset from the
Fundamental Plane (e.g., Statler et al.\ 1996; Reda et
al.\ 2005). Given this galaxy's recent (wet) merger history, its small
core size may be because of loss cone regeneration via newly produced
and/or accreted stars. That is, a pre-existing large core in this
galaxy may be partially replenished by these new stars. We do however
caution that because this core is defined by just the inner two data
points of the deconvolved surface brightness profile, i.e., within
$0\arcsec.03$, its validity can be questioned, but it
remains unlikely that they are any bigger than reported here.  \\

\noindent{\bf NGC 3640} \\

\noindent NGC 3640 is a fast rotating (Krajnovi\'c et al.\ 2013) early-type
galaxy with $M_{V} \sim -21.8$ mag and $\sigma=182$ km s$^{-1}$.
Prugniel et al.\ (1988, see also Michard \& Prugniel 2004) noted that
this galaxy is a merger in progress, possibly with a gas-poor disc
system. Not only did they find low surface brightness structures such
as shells and ripples but they also showed that the $V-R$ radial
colour profile of this galaxy gets bluer towards the inner region.
Hibbard \& Sansom (2003) however fail to detect neutral hydrogen
associated with NGC 3640. Also, Tal et al.\ (2009) recently reported
this galaxy to be highly disturbed system revealing morphological
peculiarities. Therefore, as in the case of NGC 1700, this galaxy's
undersized core and mass deficit can be interpreted as loss cone
replenishment via new star formation and/or stellar accretion not
associated with a secondary black hole (see also Krajnovi\'c et al.\
2013). It is also possible that the observed core, defined by just
 one inner data point, may not be real.\\

\noindent{\bf NGC 7785}\\

\noindent Our core-S\'ersic model fit to this elliptical galaxy light
profile yields a core size of just 5 pc, which is too small for its
absolute magnitude $M_{V}\sim -22.0$ mag and velocity dispersion
$\sigma=$ 255 km s$^{-1}$. There is no evidence for an ongoing or a
recent merging event associated with this galaxy in the
literature. Thus, its small core size as well as undersized mass
deficit suggest that one or a few minor (instead of major)
dissipationless mergers might have taken place in the absence of loss
cone refilling. Alternatively, the apparent core in this galaxy, also
defined by one data point, may be spurious.

\section{}\label{ApppB}
%\subsection{Notes on selected individual galaxies}\label{Sec91} 

This section provides a review of five galaxies (i.e., excluding NGC
4073 already discussed above) from our 26 suspected elliptical
galaxies which were shown to have a bulge+disc stellar distribution in
the literature.

\subsection{NGC 0584}

This galaxy is classified as an elliptical galaxy in the Third
Reference Catalogue, RC3 (de Vacouleurs et al.\ 1991) and as an S0 by
Sandage \& Tammann (1981). Laurikainen et al. (2010, 2011) fit a
3-component (bulge+bar+disc) model to this galaxy's $\sim$$240\arcsec$
$K_{\rm s}$-band light profile and noted that it has a large disc-like
outer envelope and a weak inner bar. However, these (bar and disc)
components were not detected in the $\sim$$80\arcsec$ $V$-band light
profile that we modelled. We do however wish to bring them to the
 attention of readers.

\subsection{NGC 4472}

The Virgo cluster galaxy NGC 4472 is classified as an elliptical
galaxy in the RC3 but as an S0 by Sandage \& Tammann
(1981). Laurikainen et al.\ (2010, 2011) identified a large-scale disc
in this galaxy which dominates the light at large radii
($R>100\arcsec)$. In contrast, NGC 4472 was considered to be an
elliptical galaxy by Kormendy et al.\ (2009) who fit a single S\'ersic
model to its light profile over $3.5\arcsec \la R \la 877 \arcsec$. We
did not detect a disc component in our $100\arcsec$ profile (see also
Ferrarese et al.\ 2006). Interestingly, however, this galaxy has the
lowest S\'ersic value ($n=3$) from our 26 suspected elliptical
galaxies (Table 2), possibly suggesting an S0 morphology. It is
classified as a slow rotator (over its inner region) by the ATLAS3D
team (Cappellari et al. 2011).

\subsection{NGC 4552}

Like NGC 4472, NGC 4552 is a member of the Virgo cluster. It is
classified as an elliptical galaxy in the RC3, but was recognized as
an S0 by Sandage \& Tammann (1981). While Kormendy et al.\ (2009)
adopted an elliptical morphology for this galaxy, their S\'ersic model
fit to its light profile over $1.3\arcsec \la R \la 495\arcsec$ shows
a clear residual structure, in agreement with the S0 morphology
adopted by Laurikainen et al.\ (2010, 2011). However, it too is
classified as a slow rotator by Cappellari et al.\ (2011) perhaps
because they only sampled the pressure-supported, bulge-dominated
inner portion of the galaxy. The disc light (e.g., Laurikainen et
al.\ 2011), dominant at large radii, does not contribute to our
$\sim$$100\arcsec$ light profile (see also Ferrarese et al.\ 2006).

\subsection{NGC 5813} 

\begin{figure*}
\includegraphics[angle=0,scale=0.68]{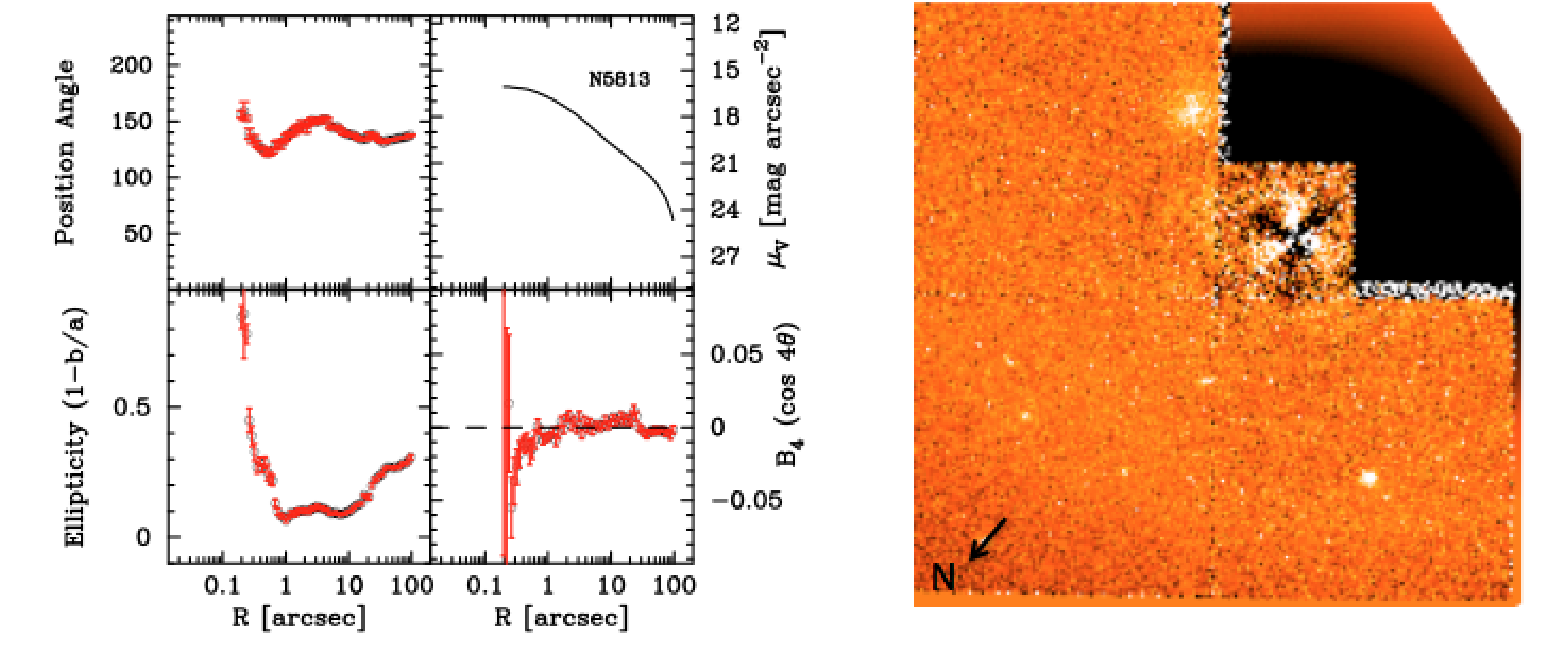}
\caption{Left panel: major-axis surface brightness, position angle,
  ellipticity and isophote shape parameter ($B_{4}$) profiles for NGC
  5813. Right panel: residual image of NGC 5813 created by subtracting
  our model image from the raw {\it HST}/WFPC2 $V$-band galaxy image.
  North is in the direction of the arrow.}
\label{FigIII3}
\end{figure*}

As hinted at in Section \ref{Sec3.2}, although NGC 5813 is classified as an
elliptical galaxy (E1-2) in RC3, we find that its ($\sim$$80\arcsec$) light
profile is best fit using the core-S\'ersic bulge+exponential model
with a small rms residual of $0.015$ mag arcsec$^{-2}$. This suggests
the galaxy may be an S0 disc galaxy, consistent with its steadily
increasing ellipticity at $R > 10\arcsec$ as well as the kinks in the
position angle and $B_{4}$ profiles at $R \sim20\arcsec$
(Fig.~\ref{FigIII3}). The residual image of this galaxy shown in
Fig.~\ref{FigIII3} is regular, there is no evidence for a distinct
morphological feature. Trujillo et al.\ (2004) also concluded that NGC
5813 may be an S0 galaxy which is better described using a bulge+disc
model. This however appears to disagree with the rotation curve given
by Efstathiou, Ellis \& Carter (1982). Using long slit data, these
authors showed that the core ($R\la 7\arcsec$) of NGC 5813 rotates
rapidly, while beyond $R\sim$$10\arcsec$ the galaxy shows little
rotation: evidence for a kinematically decoupled core.

\subsection{NGC 5419}

This galaxy is classified as an elliptical galaxy in the RC3, yet it
was identified as an S0 by Sandage \& Tammann (1981) and also recently
by Laurikainen et al.\ (2011).  We fit a core-S\'ersic bulge plus a
gaussian point source model to $\sim$$100\arcsec$ ($V$-band) light
profile, while Laurikainen et al.\ (2011) fit a three-component
(bulge+lens+outer-disc) model to a more radially extended $K_{\rm
  s}$-band brightness profile. Our $\sim$$100\arcsec$ data does not
probe this galaxy's large-scale disc. Also, the lens component is not
seen in our $V$-band light profile.

\section{}\label{ApppC}

\subsection {Light profile modelling comparison with some recent studies}

 Recently, Krajnovi\'c et al.\ (2013) fit the Nuker model
 (Kormendy et al.\ 1994) to the nuclear light profiles of 135
 ATLAS$^{\rm 3D}$ galaxies. While these authors are aware that the
 Nuker model is not a robust parametrization (i.e., the parameters are
 unstable), this model was used for the purpose of distinguishing core
 and coreless galaxies. As in the previous Nuker model works (e.g.,
 Lauer et al.\ 2005) they opt to use $\gamma'$ (the slope of the Nuker
 model at $0\arcsec.1$) as a diagnostic tool for separating their
 galaxies as core ($\gamma'\la 0.3$), intermediate (0.3 $ < \gamma' <$
 0.5) and power-law ($\gamma'\ga 0.5$) type.  However, it has been
 shown (Graham et al.\ 2003; Dullo \& Graham 2012, their Section 5.1)
 that the distance-dependent $\gamma'$ is not a physically robust or
 meaningful quantity to use. Moreover, galaxies with small S\'ersic
 indices will have $\gamma' < 0.3$ but no central deficit relative to
 the inward extrapolation of their outer S\'ersic profile.

% If we are to truly
% understand the nuclear structure of galaxies, then we must stop using
% inadequate models fit to subjective radial ranges.

In Dullo \& Graham (2012, see also Graham et al.\ 2003), we
highlighted that the Nuker model is not ideal for describing central
light profiles, and we revealed that seven ($\sim 18\%$) of the sample
galaxies with low $n$ and thus shallow inner profile slopes were
misidentified as galaxies having depleted cores by the Nuker
model. Four of these seven galaxies (NGC 4458, NGC 4473,
NGC 4478 and NGC 5576) are in common with Krajnovi\'c et al.\ (2013).
Krajnovi\'c et al.\ (2013) classify three of them (NGC 4458, NGC 4478
and NGC 5576) as an intermediate type (i.e., coreless galaxies based
on their criteria) in agreement with Dullo \& Graham (2012) and at
odds with Lauer et al.\ (2005). They, however, reminded the reader
that the Nuker model actually detected cores in all these galaxies but
this detection may change depending on the range over which the fits
were done. It is perhaps worth clarifying that the Nuker model break
radii reported by Lauer et al.\ (2005) for NGC 4458, NGC 4478 and NGC
5576 are $\sim 40 \%$ smaller than those from Krajnovi\'c et al.\
(2013). These differing break radii have resulted in a significant
difference in the $\gamma'$ measurements from these two works. For
these three galaxies in question, Krajnovi\'c et al.\ (2013) found
$\gamma'$ values that are $\sim2$ times larger than those from Lauer
et al.\ (2005). For the remaining common galaxy NGC 4473, Krajnovi\'c
et al.\ (2013) found $\gamma'=0.1$ and subsequently classified it as a
core galaxy in agreement with Lauer et al.\ (2005) and in contrast to
Ferrarese et al.\ (2006), Kormendy et al.\ (2009) and Dullo \& Graham
(2012; 2013, their section 9.2.2). NGC 4473 does not have a central
light deficit but quite the opposite: an embedded stellar disk (e.g.,
Krajnovi\'c et al.\ 2006; Capellari et al.\ 2007).
%, highlighting the
%problem with Nuker model classifications.

\label{lastpage}
\end{document}